\documentclass[AER]{AEA}
\usepackage{natbib}
\usepackage{enumitem}
\usepackage{tabularx}
\usepackage{booktabs}
\usepackage{placeins}
\usepackage{graphicx}
\usepackage{threeparttable}
\usepackage{pythonhighlight}
\usepackage[utf8]{inputenc}
\usepackage{url}
\usepackage{bibunits} 
\draftSpacing{1.5}

\newcommand{\fignote}[1]{\par\vspace{0.5ex}\begin{minipage}{\textwidth}\footnotesize\textit{Note:} #1\end{minipage}}

\begin{document}

\title{Narratives to Numbers: Large Language Models and Economic Policy Uncertainty}
\shortTitle{Narratives to Numbers}
\author{Ethan Hartley\thanks{Hartley: University of Hawaii at Manoa (email: ehartley@hawaii.edu) Acknowledgements: The technical support and advanced computing resources from University of Hawaii Information Technology Services – Cyberinfrastructure, funded in part by the National Science Foundation CC* awards \#2201428 and \#2232862 are gratefully acknowledged. The author is funded by the Department of Energy Innovation in Buildings Graduate Research Fellowship. There are no other sources of financial support or conflicts of interest to report. }}
\date{\today}
\pubMonth{11}
\pubYear{2025}
\pubVolume{Vol}
\pubIssue{Issue}

\begin{abstract}
We study the implications of adopting machine-generated data in economic analysis. Rather than taking the data-generating process as given and correcting estimates econometrically, we evaluate how upstream modeling and researcher choices shape downstream estimates. Revisiting the Economic Policy Uncertainty index, we find that large language models fine-tuned on human audit labels identify nearly twice as many auditor-flagged articles as keyword-based alternatives and are robust to noisy text and linguistic drift. Varying classifier, threshold, and aggregation choices produces substantial differences not only in performance but in what the index captures, reshaping the relationship between policy uncertainty and returns.
\end{abstract}


\maketitle

A growing share of the data in empirical economics is generated from unstructured sources via predictive modeling \citep{tetlock, gentzkow_10, gentzkow_tad, dell_as, dell_newswire}. This capacity to leverage new information is powerful, but it introduces a fundamental challenge because these generated regressors do not necessarily exhibit the classical measurement error assumed in standard econometric designs. While a rapidly developing literature focuses on valid inference with machine-learned variables \citep{battaglia2025, carlson, ludwig}, it often treats the upstream prediction model as a fixed, mechanical step. We demonstrate instead that the algorithm chosen to compress unstructured data into structured scalars shapes both the measurement error and the substantive content that ultimately enters downstream estimation.

We study these issues in the context of economic policy uncertainty (EPU). The canonical index introduced by \citet{BBD_16} (henceforth BBD) lies at the center of a large empirical literature that uses keyword-based heuristics to turn text into data and study how uncertainty shapes macroeconomic and financial outcomes \citep[e.g.,][]{thaqeb, caldara, kong, kitsul}. The policy uncertainty setting is a natural choice for two reasons. First, although dictionary rules are sensitive to domain-specific meaning and term selection \citep{loughran, keith}, keyword-based methods remain standard in this literature, creating an opportunity to assess the gains from contemporary language models. Second, the mapping from text to index is typically inherited rather than examined, even though it determines which content the index admits and how it behaves in downstream analysis. This setting allows us to study both how predictive models improve text-based measurement and how the choices underlying that measurement shape economic conclusions.

We begin by revisiting the classification step underlying the BBD index. We train fine-tuned large language models (LLMs) on BBD's original human-audited corpus, allowing us to compare contemporary text-classification methods against the keyword-based approach on a common benchmark. Our preferred model identifies nearly twice as many EPU-related articles as the keyword method, substantially expanding the coverage of policy uncertainty in newspaper text. LLMs also allow us to vary classification thresholds, aggregate predicted probabilities, and examine how alternative definitions of EPU change the content of the resulting index. We apply our preferred model to approximately 360 million historical newspaper articles from American Stories \citep{dell_as} to construct a U.S. EPU index spanning 1800--1964, extending news-based measurement of policy uncertainty a full century further into the historical record. We make the trained models and historical index publicly available.

These improvements in article-level classification raise a further question: do better predictions translate into better economic measurement? An index aggregates individual classification decisions into a measure used in downstream analysis, and changes in the underlying classifier can alter not only which articles are identified but also the content represented by the resulting index. We examine this distinction by revisiting the relationship between EPU and stock returns documented by \citet{arouri}. Whereas the canonical BBD index reproduces the negative contemporaneous relationship, our LLM-based indices do not.

To understand why the economic relationship changes, we decompose the index by article sentiment and vary the classification threshold. The negative relationship is concentrated in negatively framed coverage, while changing the threshold reveals that the relationship is sensitive to articles near the margin of the EPU definition, where policy uncertainty overlaps with war and geopolitical risk. These results suggest that differences across indices reflect not only classification error, but also the substantive content each classifier admits. A classifier can better match human judgments at the article level while producing an aggregate index that captures a different signal and yields different downstream estimates.

Our findings bridge the literature on text-based uncertainty measurement and recent work on econometric inference with unstructured data. Prior efforts to refine EPU indices augment keyword rules with sentiment weighting, co-occurrence constraints, or topic models \citep{choi, charemza_bow, azqueta}, expand keyword-based series across countries \citep{ahir, hong, diakonova}, or apply supervised classifiers and language models within specialized domains or weakly labeled samples \citep{tobback, noailly, trust, audrino, ito}. Extending recent applications of foundation models to economic prediction \citep{vafa2024career, dulabor}, we train directly on BBD's human audits to benchmark modern LLMs against the canonical dictionary rule and show that upstream modeling choices govern more than predictive accuracy. In applications that convert unstructured data into regressors, outcomes, and controls, the prediction model is often evaluated separately from the downstream analysis, while the resulting variable is treated as a fixed input. Our results show why the choices linking prediction to aggregation deserve direct attention: model architectures, classification thresholds, and aggregation rules jointly determine both the error properties of the generated variable and the economic signal it carries.

\section{From Narratives to Numbers}
As proposed in BBD, constructing a text-based EPU index proceeds in two steps. First, a bag-of-words (BOW) classifier labels an article as EPU-related if it contains at least one term from each of three keyword groups:
\begin{itemize}[noitemsep, topsep=4pt]
    \item Economic: \textit{economic} or \textit{economy}
    \item Policy: \textit{Congress}, \textit{deficit}, \textit{Federal Reserve}, \textit{legislation}, \textit{regulation}, or \textit{White House}
    \item Uncertainty: \textit{uncertain} or \textit{uncertainty}
\end{itemize}
Let $a_{ijt}$ denote the indicator that article $j$ in newspaper $i$ in period $t$ meets this criterion. Second, these labels are aggregated into a newspaper- and period-specific share,
\begin{equation}\label{eq:bbd_share}
    X_{it} = \frac{1}{J_{it}} \sum_{j=1}^{J_{it}} a_{ijt},
\end{equation}
where $J_{it} > 0$ is the total number of articles. Let $\mathcal{I}_t$ denote the set of newspapers active in period $t$, and let $T_0$ denote the normalization window. Each newspaper's series is standardized by its standard deviation over $T_0$, $Y_{it} = X_{it}/\sigma_i$; these are averaged across active newspapers, $Z_t = |\mathcal{I}_t|^{-1}\sum_{i \in \mathcal{I}_t} Y_{it}$, and rescaled to have mean 100 over $T_0$:
\begin{equation}\label{eq:bbd_index}
    \text{EPU}_t = 100 \times \frac{Z_t}{\bar{Z}}, \qquad \bar{Z} = \frac{1}{|T_0|} \sum_{t \in T_0} Z_t.
\end{equation}
Intuitively, the index rises when outlets devote more space than usual to policy uncertainty. 
 
\subsection{Index Construction with a Probabilistic Classifier}
 
Following \citet{gentzkow_tad}, we view an index as the output of a mapping from text to labels, followed by aggregation. A probabilistic classifier makes each component of this mapping explicit.
 
\medskip
\noindent\textsc{Definition 1}: \textit{An index construction consists of}
\begin{enumerate}
    \item[(i)] \textit{a scoring model that maps the representation $C_{ijt}$ of each article to a score $\hat{p}_{ijt} = f(C_{ijt}; \theta) \in [0,1]$, interpreted as the probability that the article is EPU-related, with $\theta$ estimated from labeled data or, for a dictionary, set by the researcher;}
    \item[(ii)] \textit{a threshold $\tau \in [0,1]$ that converts scores into labels, $\hat{A}_{ijt}(\tau) = \mathbf{1}\{\hat{p}_{ijt} \geq \tau\}$; and}
    \item[(iii)] \textit{an aggregation rule, either the binary share $X_{it}(\tau) = J_{it}^{-1} \sum_{j} \hat{A}_{ijt}(\tau)$ or the probabilistic share $X^{p}_{it} = J_{it}^{-1} \sum_{j} \hat{p}_{ijt}$.}
\end{enumerate}
\medskip
 
Standardization and normalization then proceed as in equation~\eqref{eq:bbd_index}. The BOW classifier is the special case in which $f$ takes only the values zero and one: the threshold plays no role, and the binary and probabilistic shares coincide. Under a keyword rule, the threshold and aggregation rule are fixed by construction; under a probabilistic classifier, each becomes a choice the researcher must make.

\subsection{Error and Content}
Definition 1 describes how an index is built, but what matters downstream is which articles it counts. For a given scoring model and threshold, each newspaper-period contributes a well-defined set of admitted articles, and two indices built from the same corpus differ only insofar as these sets differ. Characterizing that set, along with what kinds of coverage it contains, is therefore the natural starting point for comparing constructions.

\medskip
\noindent\textsc{Definition 2}: \textit{The content boundary of a binary index in newspaper $i$ and period $t$ is $B_{it}(\tau) = \{ j : \hat{p}_{ijt} \geq \tau \}$. Given a partition of articles into coverage types $k \in \mathcal{K}$, the composition of the index is the share of admitted articles of each type, $w_k(\tau) = \sum_{i,t} |B_{it}(\tau) \cap \mathcal{K}_k| \,/\, \sum_{i,t} |B_{it}(\tau)|$.}
\medskip

Let $\mathcal{A}_{it} = \{ j : A_{ijt} = 1 \}$ denote the set of articles a human auditor would label EPU-related. Because policy uncertainty is a latent concept and human auditors report varying certainty in their own judgments, $\mathcal{A}_{it}$ serves as a benchmark rather than absolute ground truth.\footnote{In the BBD audit, auditors express varying degrees of confidence, and reasonable readers often disagree on articles near the margin of the definition. Some discrepancy between $B_{it}(\tau)$ and $\mathcal{A}_{it}$ therefore reflects inherent conceptual ambiguity rather than classifier failure.} Relative to this benchmark, misclassification is the discrepancy between the two sets: false positives $B_{it}(\tau) \setminus \mathcal{A}_{it}$ create spurious spikes, and false negatives $\mathcal{A}_{it} \setminus B_{it}(\tau)$ mute true movements in the index. Standard performance metrics summarize this discrepancy. Recall, $|B_{it}(\tau) \cap \mathcal{A}_{it}| / |\mathcal{A}_{it}|$, is the share of auditor-labeled articles the index admits, and precision, $|B_{it}(\tau) \cap \mathcal{A}_{it}| / |B_{it}(\tau)|$, is the share of admitted articles that auditors would label EPU-related.

A baseline characterization treats discrepancies across indices as classical measurement error: each construction $m$ tracks the same homogeneous latent process, $I^m_t = U_t + e^m_t$, where the noise $e^m_t$ arises from false positives and false negatives and is uncorrelated with both $U_t$ and the regression error. Under this view, indices differ only in how much noise they carry, and a better classifier matters downstream solely by reducing bias in the articles identified as EPU. 

However, constructions can also differ in the substance of \textit{what} they measure. Coverage of a looming war, potential tariffs, and a routine budget debate, or coverage framed with alarm versus reported matter-of-factly, may all qualify as EPU-related, yet relate quite differently to an economic outcome such as stock returns. If type-$k$ coverage carries a distinct effect $\beta_k$, regressing an outcome on index $m$ heuristically recovers a weighted average, $\beta^m \approx \sum_k w^m_k \beta_k$, where the effective weights reflect the index's composition $w^m_k$ and each type's contribution to time-series variation in the index. Two indices will therefore yield different downstream estimates whenever coverage effects are heterogeneous ($\beta_k \neq \beta_{k'}$) and the constructions weight those coverage types differently ($w^m_k \neq w^{m'}_k$).

These two channels, classification error and substantive composition, are governed jointly by the components of Definition 1. A more expressive scoring model reduces error relative to $\mathcal{A}_{it}$, but scoring models also differ systematically in which coverage types they admit: a keyword rule, for example, captures only articles containing specific surface level cues, whereas a contextual model admits a broader set of policy discourse. Holding the scoring model fixed, the decision threshold $\tau$ and aggregation rule shift $B_{it}(\tau)$ and $w_k(\tau)$ directly. Raising $\tau$ isolates articles near the unambiguous core of the EPU definition at the cost of a smaller sample, while lowering $\tau$ admits more marginal coverage, such as geopolitical or military events intersecting with economic policy, while expanding the pool of positive articles.\footnote{In data-scarce settings, such as historical corpora or disaggregated sub-indices, a stringent threshold can leave too few positive articles per cell to form a stable series.} Because indices with different compositions capture different economic signals, they are not simply noisier versions of one another, and regressions that use them may identify correspondingly different relationships.
\label{sec1}

\section{LLMs for EPU}\label{sec2}
LLMs are large neural networks pre-trained on broad text using simple predictive targets (for example, next-word prediction). Trained on vast corpora, they learn general language representations that transfer easily and can be adapted to new tasks with limited labeled data. Recent advances, most notably the Transformer and large-scale pretraining of LLMs, have made previously infeasible text-as-data tasks routine \citep{vaswani, devlin, brown,touvron}. As \citet{gentzkow_tad} emphasize, the high dimensionality and unstructured nature of text have long posed a barrier to empirical work; LLMs address this through attention. Rather than building dependence through fixed local structure or recursion, attention forms a data-driven weighted average over the full context. For token \(t\), the model outputs \(y_t=\sum_{s=1}^{T}\alpha_{ts}v_s\), where the weights \(\alpha_{ts}\) are nonnegative and sum to one (a learned, content-based weighting). Positional encodings supply order, and fully parallel computation enables scale. The result is a compact, context-aware representation of high-dimensional text that embeds each document in its broader linguistic context. 

LLMs can be deployed for empirical work in several ways. In \emph{language modeling}, the objective is next-token prediction, useful for text generation or completion \citep{vafa2024career, dulabor}. In \emph{instruction use}, an instruction-tuned model is queried with structured prompts to return labels or schema-constrained outputs, useful when labeled data are scarce \citep{brown, gilardi} but vulnerable to prompt sensitivity, hallucinated content, and calibration concerns \citep{lin, zhao, ullman}. In \emph{supervised adaptation}, a pre-trained backbone is paired with a small task head trained on labeled data, allowing the empiricist to steer predictions toward domain-relevant cues, align output with human labels, and obtain calibrated probabilities while eliminating prompt sensitivity. We use supervised adaptation as our primary deployment regime.

We combine an LLM backbone $\mathcal{M}_{\text{LLM}}$ with a classification head $\mathcal{M}_{\text{CH}}$.
Given raw text $\mathbf{C}$, the backbone produces a task-relevant representation
\[
\tilde{\mathbf{C}} \;=\; \mathcal{M}_{\text{LLM}}(\mathbf{C};\, \theta_{\text{LLM}}).
\]
The head $\mathcal{M}_{\text{CH}}$ then maps $\tilde{\mathbf{C}}$ to a probability of the positive class (e.g., “article reflects policy uncertainty”) via a logistic link:
\[
p(y{=}1 \mid \mathbf{C}) \;=\; \sigma\!\big( \mathbf{w}^\top \tilde{\mathbf{C}} + b \big),
\]
where $\sigma(z)=1/(1+e^{-z})$ and $(\mathbf{w},b)$ are the head parameters. Econometrically, this is a logit model on the learned representation $\tilde{\mathbf{C}}$. Fine-tuning updates $\mathcal{M}_{\text{CH}}$ and (optionally) the backbone parameters $\theta_{\text{LLM}}$ to minimize a supervised loss, typically cross-entropy. In our main specifications we fine-tune both the backbone and the classification head. Exercises that tune only the classification head are reported in the Supplemental Appendix.

\subsection{Data}
A central component of BBD is their comprehensive human audit study in which teams of researchers systematically reviewed and labeled approximately 12,000 news articles for their relation to policy uncertainty. The resulting dataset covers the period 1900-2012 and provides binary labels for EPU, supplemented by auditor certainty scores. From this corpus, we remove duplicate entries and articles without recoverable text, leaving 10,393 articles with human-designated labels. We randomly split the data into 70\% training (n = 7,275), 20\% validation (n = 2,078), and 10\% test (n = 1,040). As a robustness check, we also use a temporal split: articles from 1900-2005 go to training (7,660) and validation (1,915), and articles from 2006-2012 form the held-out test set (818). We update model weights on the training set, tune model hyperparameters on the validation set, and report final performance on the held-out test set to limit overfitting.

\subsection{Methods}
Our goal is to improve the mapping $\mathcal{M}$ from text to numerical labels, prioritizing predictive performance and out-of-sample generalizability while acknowledging practical constraints common in applied settings. We evaluate three open-source LLMs to quantify trade-offs in predictive performance, speed, and input flexibility: BERT \citep{devlin}, Longformer \citep{beltagy}, and Llama 3.1 (8B) \citep{llama3}. BERT is a compact encoder that trains quickly on standard hardware but has a short context window. Longformer is designed for efficient long-document processing via sliding-window attention. Llama 3.1 (8B) is a decoder LLM with stronger contextual representations and multilingual capacity at higher computational cost.

We fine-tune each architecture at maximum input lengths of 32 (titles only), 128, 256, 512, 1024, and 2048 tokens. For each specification, we select hyperparameters via an optimized search over validation loss using Optuna \citep{optuna}. Results from BERT, which performs consistently below the other two models, are omitted from the main text but reported in the Supplemental Appendix. We therefore focus on Longformer and Llama 3.1. Longformer is lighter and more practical for researchers with limited compute; Llama 3.1 offers multilingual support and stronger performance at greater computational cost. 

All BOW classifiers use the term sets established by BBD. As a traditional machine-learning benchmark, we also train support vector machines (SVMs) on unigram features, a common non-dictionary alternative in the text-as-data literature \citep[e.g.,][]{tobback}. Unless otherwise noted, the keyword rule and the SVM both use the full text of each article. We release all selected models on the HuggingFace Hub and provide full training details, hyperparameter search spaces, and example code in the Supplemental Appendix.

\FloatBarrier 
\subsection{Classifier Performance}
\begin{figure}[!hbtp]
      \centering
      \includegraphics[width=\textwidth]{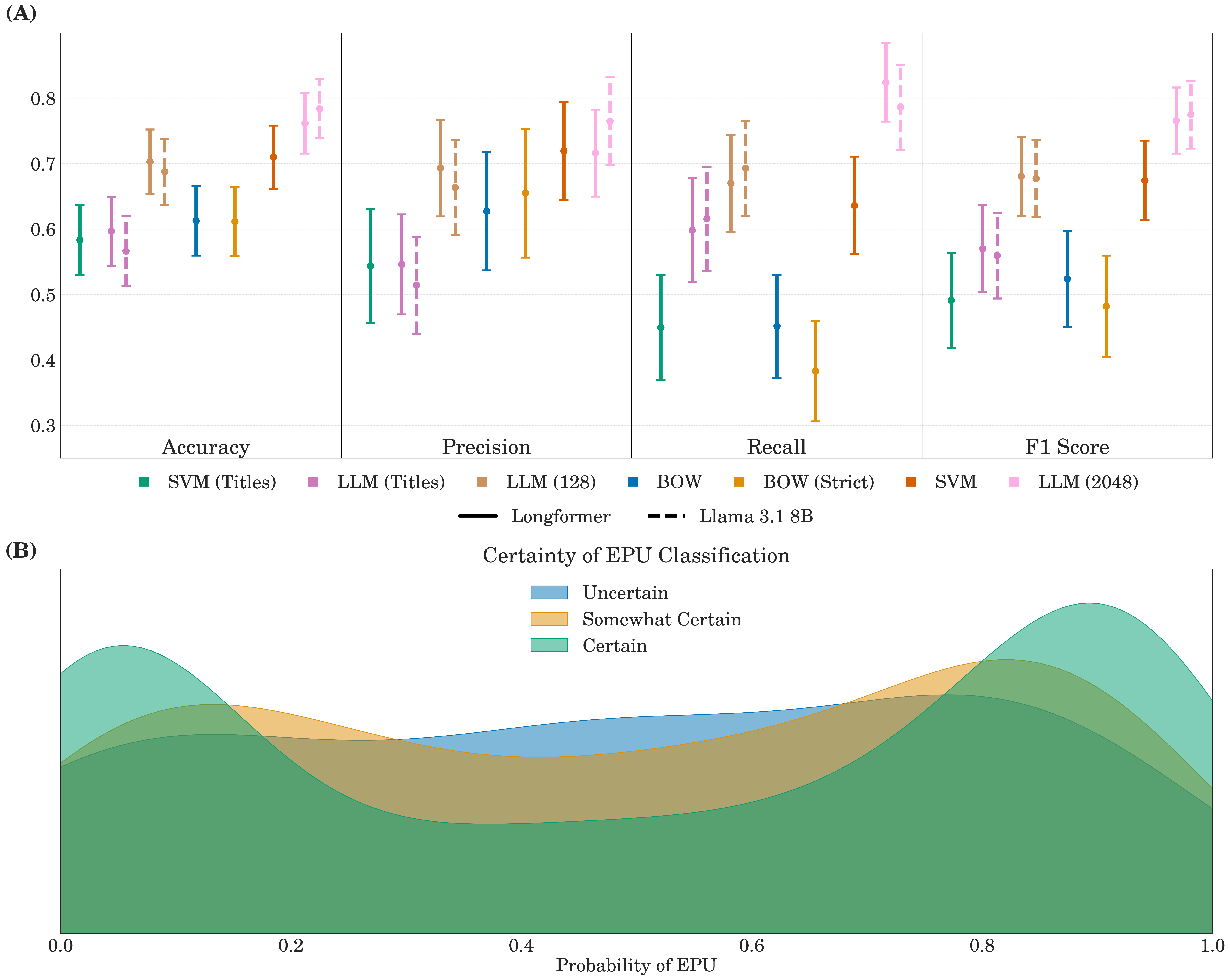}
      \caption{Article-Level Classification Performance}
      \label{fig:performance}
      \begin{figurenotes}
          Panel A reports accuracy, precision, recall, and $F_1$ for the BOW, SVM, and LLM classifiers; LLMs use article titles, the first 128 tokens, or the first 2,048 tokens. Solid lines denote Longformer and dashed lines Llama 3.1 (8B). Points are bootstrap means with 95 percent confidence intervals. LLMs are evaluated on the held-out test set at the Youden threshold, and BOW on the full audit corpus. Panel B plots kernel densities of Longformer (2048) scores $\hat{p}_{ijt}$ on held-out test articles, by auditor-reported certainty.
      \end{figurenotes}
  \end{figure}

We evaluate each classifier against the BBD audit labels using accuracy, precision, recall, and their harmonic mean ($F_{1}$). Probabilistic models return continuous scores $\hat{p}_{ijt} \in [0,1]$, which we convert into binary labels using decision thresholds tuned on the validation set. In Figure \ref{fig:performance}A, we select the threshold that maximizes Youden's index \citep{youden},
\[
\tau^* = \arg\max_{\tau} \{\operatorname{TPR}(\tau) - \operatorname{FPR}(\tau)\},
\]
where the true positive rate $\operatorname{TPR}(\tau)$ is recall and the false positive rate $\operatorname{FPR}(\tau)$ is the share of auditor-negative articles the classifier admits. This criterion equally weights sensitivity and specificity, balancing true-positive and true-negative rates. Unless noted otherwise, we report performance for all trained models on the held-out test set and evaluate BOW on the full audit corpus.

Figure \ref{fig:performance}A reports bootstrapped performance for the baseline BOW classifier, an SVM, and several LLM configurations. Even with minimal text, contextual models match or beat keyword rules: LLMs trained on article titles alone match the $F_1$ score of BOW applied to full articles, and using just the first 128 tokens exceeds BOW and matches full-text SVMs. Extending the window to 2048 tokens delivers a 46 percent relative $F_1$ improvement over BOW. Most notably, LLMs uniformly improve recall, raising it from 45 percent under BOW to 82 percent for the best model. We use Longformer (2048) as our baseline specification, as it matches the performance of Llama 3.1 while remaining fast enough to scale to millions of articles.\footnote{The Supplemental Appendix reports additional diagnostics across alternative sample splits, sequence lengths, and multi-label policy category predictions.}

Beyond improving binary classification, a good scoring model should produce probabilities $\hat{p}_{ijt}$ that reflect the strength of the underlying signal. A natural concern with heavily parameterized models is that their predictions can be opaque, prompting a growing literature on LLM interpretability \citep{shap,cunningham,dunefsky}. Rather than opening the black box mechanically, we ask a more direct question for economic measurement: do the model's scores mirror human judgment? Policy uncertainty is a latent concept with ambiguous boundaries, so a model that has learned the concept should assign extreme scores to clear-cut articles and intermediate scores to marginal ones where human auditors are unsure. 

Figure \ref{fig:performance}B confirms this pattern, plotting kernel density estimates of $\hat{p}_{ijt}$ from Longformer (2048) across levels of auditor-reported certainty on the held-out test set. As auditor certainty rises, probability mass shifts away from 0.5 toward the tails, reflecting more decisive model assessments on clearer texts. Correspondingly, while the BOW misclassification rate remains flat at roughly 40 percent across all certainty levels, Longformer's error rate falls from 37 percent on ``uncertain'' articles to 18 percent on ``certain'' articles, a 51 percent relative reduction. Given that the index is meant to proxy how news shapes perceived policy uncertainty, it is reassuring that the mapping from text to scores moves with human certainty, assigning more extreme probabilities to clearer, less ambiguous articles.


\section{From Better Classifiers to Better Indices}\label{sec3}
For economists, article-level classification is only a means to an end: what enters regressions, forecasts, and policy debates are the aggregate indices built from those predictions. This section examines how article-level gains translate into index-level measurement. We show that LLM-based indices track human audit benchmarks far more closely than keyword rules, that threshold and aggregation choices become consequential at scale, and that contextual models extend measurement to noisy historical archives where dictionary methods fail.

\FloatBarrier 
\subsection{Index Fidelity, Thresholds, and Aggregation}\label{index_error}
\begin{figure}[!hbtp]
      \centering
      \includegraphics[width=\textwidth]{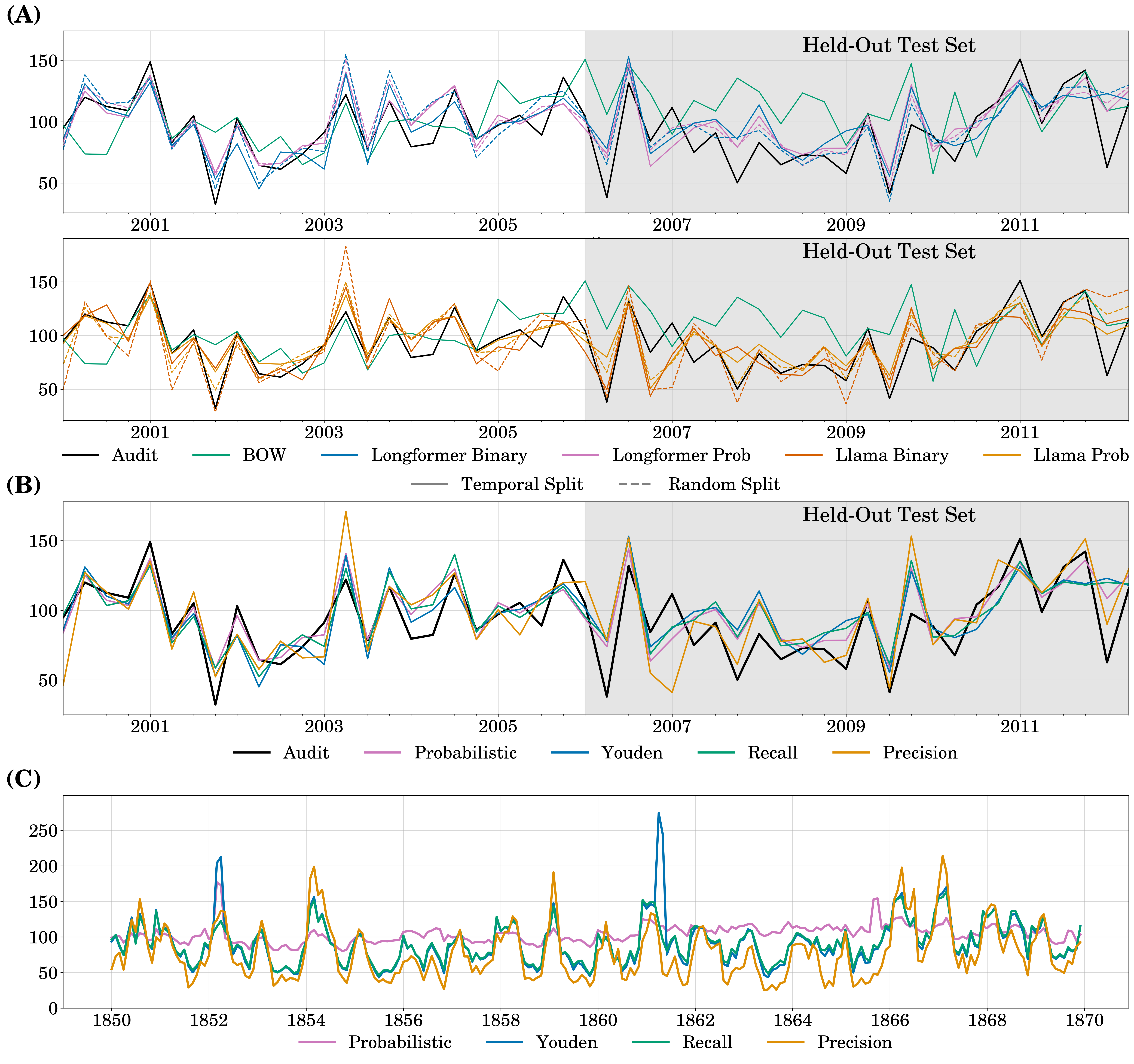}
      \caption{From Article-Level Performance to Index-Level Measurement}
      \label{fig:index_comparison}
      \begin{figurenotes}
          Panels A and B show monthly indices built from the BBD audit sample; the black line is the human-audit benchmark and shading marks the 2006--2012 held-out period. Panel A compares BOW, Longformer, and Llama indices under binary and probabilistic aggregation; solid lines use the temporal split and dashed lines the random split. Over the held-out period, correlations with the benchmark are 0.24 for BOW, 0.70 (binary) and 0.78 (probabilistic) for Longformer, and 0.71 and 0.71 for Llama. Panel B holds Longformer fixed and compares probabilistic aggregation with the Youden, recall, and precision thresholds. Panel C applies the four Panel B constructions to 28.4 million American Stories articles, 1850--1870.
      \end{figurenotes}
  \end{figure}

A natural first question is whether lower article-level error actually yields an index that tracks the human audit benchmark more closely. In the framework of Section I.B, reducing false positives and false negatives shrinks the measurement error $e^m_t$ in $I^m_t = U_t + e^m_t$. Evaluating an aggregate time series requires a contiguous out-of-sample window, since a random split mixes training and test articles within every month. Figure~\ref{fig:index_comparison}A therefore evaluates the models under the temporal split, which reserves all articles from 2006--2012 as a clean held-out period, and compares each model's index against the benchmark series constructed directly from human auditor labels $\mathcal{A}_{it}$. Across both binary and probabilistic constructions, LLM-based indices track the human benchmark far more closely than the BOW index.

Once a researcher moves from a deterministic keyword rule to a probabilistic classifier, constructing an index requires choosing a decision threshold $\tau$ and an aggregation rule (Definition 1). Figure~\ref{fig:index_comparison}B illustrates how these choices behave within the audit sample using four constructions of our baseline random-split Longformer (2048) model. All thresholds are calibrated once on the validation split of the audit corpus and are never re-tuned on the test set, the historical corpus, or any downstream outcome; the indices in Figures~\ref{fig:index_comparison} and \ref{fig:historical} and Table~\ref{tab:arouri} all use these fixed thresholds. Specifically, the Youden index maximizes Youden's $J$ ($\tau = 0.45$), while the high-recall ($\tau = 0.40$) and high-precision ($\tau = 0.74$) indices select the thresholds whose validation recall and precision are closest to an ex ante target of 0.85. Each threshold maps scores into the content boundary $B_{it}(\tau)$ in a transparent way. The high-recall threshold lowers the bar to capture more true EPU coverage at the cost of a higher baseline of false positives, the high-precision threshold isolates cleaner spikes around unambiguous episodes while missing more marginal coverage, and Youden's rule balances the two.

Within the small audit sample, these constructions behave similarly, and the probabilistic index ($X^p_{it}$), which sums predicted probabilities rather than binary labels, achieves the highest correlation with the human benchmark (0.81). However, Figure~\ref{fig:index_comparison}C shows that when scaled to a large corpus, threshold and aggregation choices generate markedly different series. This divergence is especially pronounced for probabilistic aggregation, which can both mute and distort the signal at scale. Small background probabilities on the vast majority of non-EPU articles accumulate to dilute true policy spikes, while broad shifts in media volume or baseline assignment probabilities can move the aggregate series even without commensurate policy news.\footnote{Looking ahead, probabilistic aggregation may become more attractive if future audits provide graded or ``soft'' labels reflecting uncertainty intensity rather than binary indicators, in line with \citet{epping2025harnessing}.} For large-scale applications, binary thresholding therefore provides a more disciplined filter, allowing the researcher to govern the trade-off between signal purity and sample coverage explicitly through $\tau$.

\FloatBarrier 
\subsection{Historical Texts at Scale}\label{hist_index}
Large-scale digitization initiatives such as American Stories apply optical character recognition (OCR) to historical newspaper archives, opening periods that lack clean administrative data to quantitative research \citep{dell_as, dell_newswire}. A central limitation of OCR, however, is that it leaves digitized text riddled with transcription errors. Keyword-based methods, which depend on exact string matches, are especially brittle in this environment; when characters are dropped, substituted, or split unexpectedly, a dictionary rule fails to match target terms and suffers severe false negatives.

LLMs, by contrast, recover meaning from noisy text by exploiting contextual patterns rather than exact spellings. We demonstrate this capacity by applying our fine-tuned Longformer (2048) model to approximately 360 million newspaper articles from the American Stories project \citep{dell_as}. Depending on the threshold $\tau$, our model flags between 1 and 3 percent of articles as EPU-related, roughly 300 to 1,000 times the 0.003 percent captured by the BBD keywords. By pooling thousands of newspapers and overcoming transcription noise, the resulting 1800--1964 series extends news-based measurement a full century earlier than existing U.S. indices and yields, to our knowledge, the first monthly policy uncertainty index for the nineteenth century. Over the 1900--1964 overlap, our series correlates with the original historical BBD index at 0.33.

\begin{figure}[!b]
      \centering
      \includegraphics[width=\textwidth]{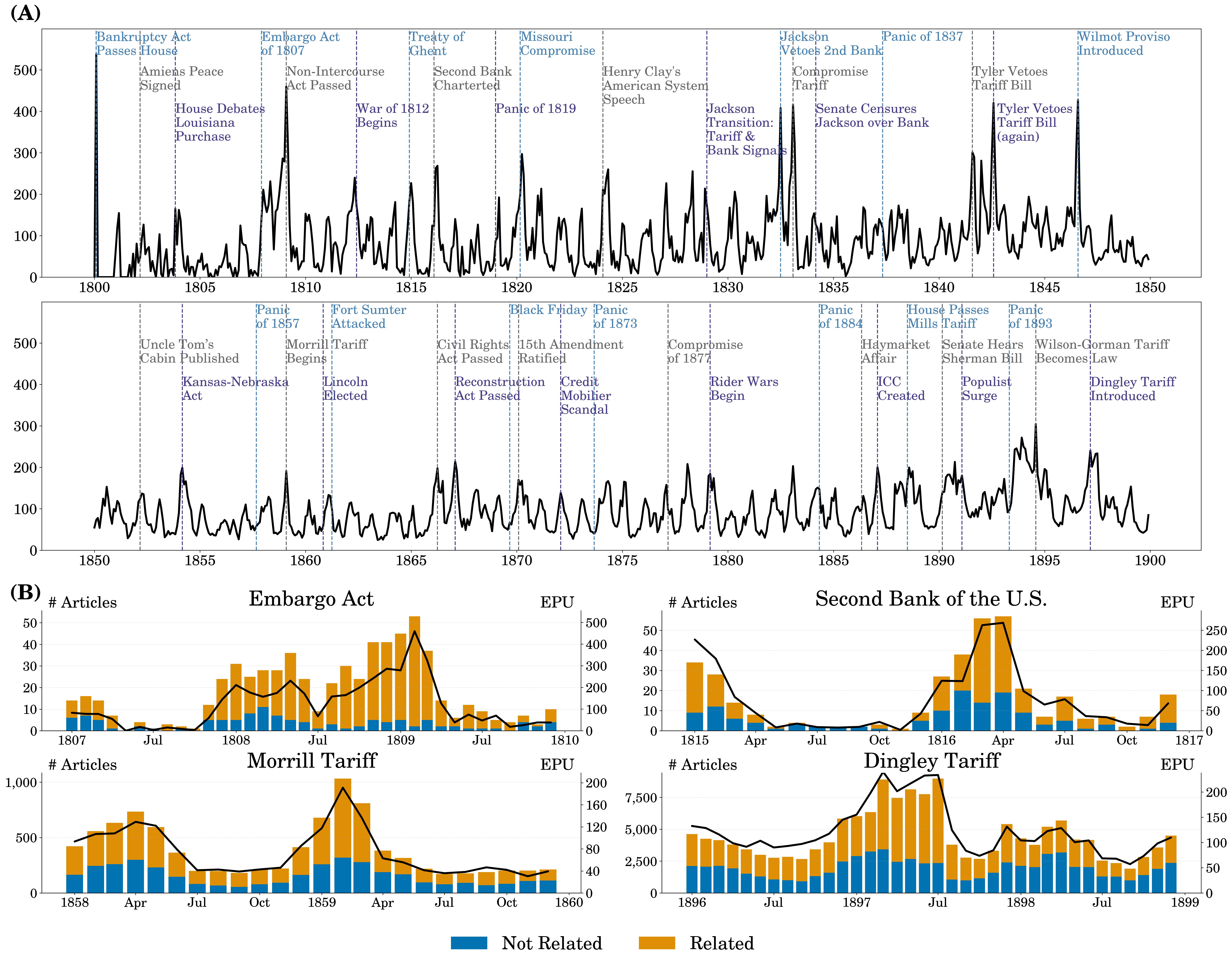}
      \caption{Historical U.S. Economic Policy Uncertainty Index, 1800--1900}
      \label{fig:historical}
      \begin{figurenotes}
          Panel A shows the monthly index from American Stories, built with Longformer (2048) at the precision threshold ($\tau = 0.74$); vertical lines mark wars, financial crises, and policy episodes. Panel B shows four episodes. Bars count EPU-flagged articles that Llama 3.1 (8B) classifies, zero-shot, as related or not related to the named event (left axis); the line is the index (right axis).
      \end{figurenotes}
\end{figure}

Figure~\ref{fig:historical}A presents the resulting index for 1800--1900 using the precision-optimized threshold ($\tau = 0.74$) to generate binary classifications. This threshold restricts $B_{it}(\tau)$ to clearly relevant articles and mitigates spurious spikes in noisy historical text. While we annotate several historical episodes that plausibly drive the observed surges, matching time-series peaks to events ex post remains an inherently ad hoc exercise. Figure~\ref{fig:historical}B illustrates how LLMs can make this validation step more systematic by directly inspecting the composition $w_k(\tau)$ of admitted articles during major spikes. Rather than assigning events to peaks purely by hand, we query Llama 3.1 (8B) Instruction in a zero-shot setting with the prompt \texttt{"Does the article relate to the topic? Topic: \{topic\} Article: \{article\}?"}. Across four illustrative episodes, the resulting classifications show that coverage of the named event dominates the admitted set during each spike, providing a direct check on the content driving elevated policy uncertainty. \enlargethispage{1\baselineskip}

\section{Revisiting Policy Uncertainty and Stock Returns}\label{sec4}
We now examine whether upstream measurement choices carry through to downstream empirical conclusions by revisiting the canonical link between policy uncertainty and stock returns. Discrepancies in downstream regressions can reflect either differences in classification error $e^m_t$ or shifts in the underlying economic signal governed by composition $w^m_k$. To separate these mechanisms, we first re-estimate the specification of \cite{arouri} using the BBD index and our historical LLM indices, then decompose the LLM indices by article sentiment, and finally examine how varying the classification threshold changes their topical composition.

\subsection{Baseline Estimates}
To test whether upstream measurement choices alter downstream conclusions, we begin by replicating the baseline linear specification of \citet{arouri}. We compare four series over 1900--1950: the published BBD benchmark and our three LLM-based historical indices from Section~\ref{hist_index}, corresponding to the recall-, Youden-, and precision-optimized thresholds from Section~\ref{index_error}.\footnote{Given the sparsity of keyword matches in \textit{American Stories} documented in Section~\ref{hist_index}, we use the published BBD index as our keyword-based benchmark.} Specifically, we estimate
\begin{equation} \label{eq:arouri}
R_{t} = \beta_{0} + \beta_{1}R_{t-1} + \beta_{2}\Delta EPU_{t} + \beta_{3}\Delta EPU_{t-1} + \epsilon_{t},
\end{equation}
where $R_{t}$ is the monthly log return on the stock market index from \citet{shiller}, $\Delta EPU_{t}$ is the monthly log change in the EPU index, and standard errors are HAC-robust with six lags \citep{newey}.

Panel A of Table~\ref{tab:arouri} reports the results. The BBD index reproduces the central finding of \citet{arouri}: a rise in policy uncertainty coincides with a significant decline in returns ($-0.033$, against their $-0.029$). Across all three LLM-based indices, however, this contemporaneous relationship vanishes. At every threshold, the contemporaneous coefficient is near zero and insignificant, while lagged uncertainty instead enters positively and significantly, a pattern absent from both BBD and the original estimates. Viewed through the $I_t^m = U_t + e_t^m$ decomposition, this divergence is revealing. If the indices simply measured a common latent process $U_t$ with different classification error $e_t^m$, then either replacing BBD with a model that better matches human auditors or raising the threshold to purge false positives should reduce $e_t^m$ and preserve the baseline relationship. That the contemporaneous effect disappears and remains unchanged across stricter thresholds implies that the discrepancy is not driven by $e_t^m$ alone. Instead, the classifiers admit different types of coverage, altering the substantive signal $U_t$ that enters the regression.

\subsection{Sentiment Decomposition}
Why indices disciplined by the same human audits capture different coverage is not immediately obvious, but differences in model capacity offer a natural explanation. Because keyword-based heuristics rely on a small set of surface lexical cues, any textual attribute correlated with those terms, such as negative tone or alarm, is mechanically bundled into the index. A contextual classifier, by contrast, fits human judgments through a broader representation that recognizes policy uncertainty whether it is framed with alarm or reported matter-of-factly. Under this view, article sentiment is left free to vary in the LLM-based indices.

\begin{table}[!htbp]\centering
\caption{Market Returns and EPU Index Composition}
\label{tab:arouri}
\begin{threeparttable}
\small
\setlength{\tabcolsep}{6pt}
\begin{tabular}{lccccc}
\toprule
& Arouri et al. & BBD & LLM: Recall & LLM: Youden & LLM: Precision \\
\midrule
\multicolumn{6}{l}{\textbf{Panel A: Baseline}}\\
\midrule
Constant   & 0.003$^{***}$ & 0.003$^{*}$ & 0.003$^{*}$ & 0.003$^{*}$ & 0.003$^{*}$ \\
         & (0.001)       & (0.002) & (0.002) & (0.002) & (0.002) \\
$R_{t-1}$  & 0.240$^{***}$ & 0.279$^{***}$ & 0.286$^{***}$ & 0.286$^{***}$ & 0.285$^{***}$ \\
         & (0.026)       & (0.044) & (0.040) & (0.040) & (0.040) \\
$\Delta$EPU$_{t}$  & -0.029$^{***}$ & -0.033$^{***}$ & -0.004 & -0.004 & -0.004 \\
         & (0.004)       & (0.012) & (0.017) & (0.016) & (0.010) \\
$\Delta$EPU$_{t-1}$ & -0.008$^{*}$ & 0.008 & 0.033$^{***}$ & 0.033$^{***}$ & 0.020$^{**}$ \\
         & (0.005)      & (0.011) & (0.010) & (0.010) & (0.008) \\
\midrule
Adjusted R$^2$   & 0.097 & 0.108 & 0.085 & 0.087 & 0.084 \\
EPU Article Count (M) & -- & -- & 6.81 & 5.82 & 1.73 \\
\addlinespace[2pt]

\midrule
\multicolumn{6}{l}{\textbf{Panel B: Negative Sentiment}}\\
\midrule
Constant   & 0.003$^{***}$ & 0.003$^{*}$ & 0.003$^{*}$ & 0.003$^{*}$ & 0.003$^{*}$ \\
         & (0.001)       & (0.002) & (0.002) & (0.002) & (0.002) \\
$R_{t-1}$  & 0.240$^{***}$ & 0.279$^{***}$ & 0.294$^{***}$ & 0.292$^{***}$ & 0.285$^{***}$ \\
         & (0.026)       & (0.044) & (0.040) & (0.040) & (0.040) \\
$\Delta$EPU$_{t}$  & -0.029$^{***}$ & -0.033$^{***}$ & -0.029$^{**}$ & -0.026$^{**}$ & -0.010 \\
         & (0.004)       & (0.012) & (0.013) & (0.012) & (0.007) \\
$\Delta$EPU$_{t-1}$ & -0.008$^{*}$ & 0.008 & 0.025$^{**}$ & 0.023$^{**}$ & 0.003 \\
         & (0.005)      & (0.011) & (0.011) & (0.010) & (0.008) \\
\midrule
Adjusted R$^2$   & 0.097 & 0.108 & 0.099 & 0.098 & 0.080 \\
EPU Article Count (M) & -- & -- & 0.85 & 0.74 & 0.26 \\
Sentiment Share (\%) & -- & -- & 12.51\% & 12.75\% & 15.22\% \\
\addlinespace[2pt]

\midrule
\multicolumn{6}{l}{\textbf{Panel C: Neutral Sentiment}}\\
\midrule
Constant   & 0.003$^{***}$ & 0.003$^{*}$ & 0.003$^{*}$ & 0.003$^{*}$ & 0.003$^{*}$ \\
         & (0.001)       & (0.002) & (0.002) & (0.002) & (0.002) \\
$R_{t-1}$  & 0.240$^{***}$ & 0.279$^{***}$ & 0.284$^{***}$ & 0.285$^{***}$ & 0.284$^{***}$ \\
         & (0.026)       & (0.044) & (0.040) & (0.040) & (0.040) \\
$\Delta$EPU$_{t}$  & -0.029$^{***}$ & -0.033$^{***}$ & 0.001 & 0.002 & -0.001 \\
         & (0.004)       & (0.012) & (0.016) & (0.015) & (0.010) \\
$\Delta$EPU$_{t-1}$ & -0.008$^{*}$ & 0.008 & 0.028$^{***}$ & 0.028$^{***}$ & 0.019$^{***}$ \\
         & (0.005)      & (0.011) & (0.009) & (0.009) & (0.007) \\
\midrule
Adjusted R$^2$   & 0.097 & 0.108 & 0.083 & 0.084 & 0.084 \\
EPU Article Count (M) & -- & -- & 5.73 & 4.89 & 1.42 \\
Sentiment Share (\%) & -- & -- & 84.16\% & 83.99\% & 81.86\% \\
\addlinespace[2pt]
\midrule
\multicolumn{6}{l}{\textbf{Panel D: Positive Sentiment}}\\
\midrule
Constant   & 0.003$^{***}$ & 0.003$^{*}$ & 0.003$^{*}$ & 0.003$^{*}$ & 0.003$^{*}$ \\
         & (0.001)       & (0.002) & (0.002) & (0.002) & (0.002) \\
$R_{t-1}$  & 0.240$^{***}$ & 0.279$^{***}$ & 0.283$^{***}$ & 0.284$^{***}$ & 0.285$^{***}$ \\
         & (0.026)       & (0.044) & (0.040) & (0.041) & (0.040) \\
$\Delta$EPU$_{t}$  & -0.029$^{***}$ & -0.033$^{***}$ & 0.011 & 0.008 & 0.002 \\
         & (0.004)       & (0.012) & (0.011) & (0.010) & (0.007) \\
$\Delta$EPU$_{t-1}$ & -0.008$^{*}$ & 0.008 & 0.014 & 0.010 & 0.008 \\
         & (0.005)      & (0.011) & (0.009) & (0.008) & (0.005) \\
\midrule
Adjusted R$^2$   & 0.097 & 0.108 & 0.081 & 0.079 & 0.081 \\
EPU Article Count (M) & -- & -- & 0.23 & 0.19 & 0.05 \\
Sentiment Share (\%) & -- & -- & 3.33\% & 3.27\% & 2.92\% \\
\addlinespace[2pt]

\bottomrule
\end{tabular}

\begin{tablenotes}[flushleft]
\footnotesize
\item \textit{Notes:} The dependent variable is the monthly log stock return. The Arouri et al.\ column reports \citet{arouri} (Table 2, column 2). All other columns use 1900--1950 ($N = 610$ months) and HAC standard errors (6 lags). LLM columns use $\tau = 0.40$, 0.45, and 0.74; Panels B--D use articles of that tone only. \textit{Sentiment Share} is the tone's share of flagged articles. *** $p<0.01$, ** $p<0.05$, * $p<0.1$.
\end{tablenotes}
\end{threeparttable}
\end{table}
\clearpage

\noindent If market returns respond primarily to coverage that conveys concern, pooling across neutral and negative articles will attenuate the estimated effect. We test this hypothesis by decomposing each LLM-based index by sentiment.

We classify article sentiment using FinBERT \citep{araci}, an off-the-shelf language model fine-tuned to label financial text as negative, neutral, or positive. We apply FinBERT without further training, so sentiment is measured independently of both the EPU classifier and the returns data.\footnote{Because FinBERT reads at most 512 tokens, we split the first 2,048 tokens of each EPU-flagged article into overlapping 512-token windows and classify each window. An article is labeled negative if at least one window is negative and none is positive, positive in the reverse case, and neutral otherwise.} For each threshold, we then rebuild separate indices by sentiment using the same construction as the full index. Across thresholds, more than 80 percent of flagged articles are neutral, 12.5 to 15.2 percent are negative, and roughly 3 percent are positive.

Panel B of Table~\ref{tab:arouri} shows how decomposing the index unpacks the pooled estimates in Panel A. In negative-sentiment coverage, higher policy uncertainty coincides with lower returns: the coefficients are $-0.029$ at the recall threshold and $-0.026$ at Youden, in line with BBD and \citet{arouri} and significantly different from their neutral counterparts ($p = 0.023$ and $p = 0.024$). Although the historical BBD index and our series draw from different newspaper archives, a pure corpus effect would not explain why isolating negative-sentiment articles within \emph{American Stories} so closely reproduces the BBD estimates. Neutral coverage shows no contemporaneous effect and instead drives the positive lag. The pooled LLM index thus loses the baseline relationship not because it is noisier, but because a more holistic classifier captures a broader range of EPU-related coverage. While a keyword rule isolates a narrow slice of text, the LLM admits the wider body of policy discourse that human auditors label as EPU, pooling acute, market-moving shocks with routine policy coverage that relates to returns differently. Even within negative coverage, however, the relationship weakens at the precision threshold, pointing to a second compositional channel: how the decision threshold itself reshapes the topical content of the index.

\subsection{Index Composition and Threshold Choice}
While Section~\ref{sec3} evaluates the classification threshold as a trade-off between classification errors, the returns exercise demonstrates that varying $\tau$ also shifts the content boundary $B_{it}(\tau)$ in an economically meaningful way. An article's score reflects both signal clarity and conceptual proximity to the core definition of policy uncertainty, so raising $\tau$ changes which topics enter the generated regressor. Thresholding therefore alters the latent signal $U_t$ rather than just the measurement error $e_t^m$, inducing compositional shifts that downstream error-correction techniques do not necessarily resolve. To quantify this filtration, we track the topical composition of the negative-sentiment index as $\tau$ increases from 0.40 to 0.90, using the economic, military, and foreign-nation categories of the Harvard General Inquirer \citep{stone} and the stock-market terms of \cite{baker_2019}.

\begin{figure}[!htbp]
      \centering
      \includegraphics[width=0.95\textwidth]{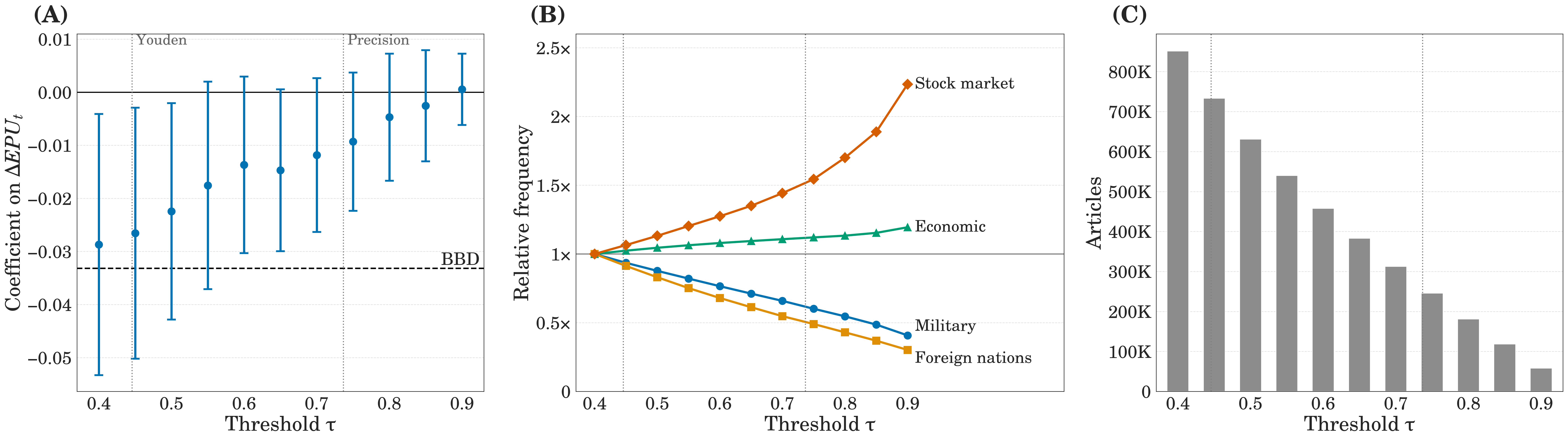}
      \caption{Threshold Choice and the Negative-Sentiment Index}
      \label{fig:content_sweep}
      \begin{figurenotes}
          Negative-sentiment index, 1900--1950. Panel A plots the coefficient on $\Delta$EPU$_t$ from equation (3) at each threshold $\tau$, with 95 percent HAC confidence intervals; the dashed line is the BBD estimate. Panel B plots dictionary term rates per 1,000 words among admitted articles, relative to $\tau = 0.40$. Panel C reports admitted articles. Dotted lines mark the Youden and precision thresholds.
      \end{figurenotes}
  \end{figure}

Figure~\ref{fig:content_sweep} traces how varying $\tau$ reshapes both the regression estimates and the underlying text of the negative-sentiment index. In Panel A, we re-estimate equation (3) at each threshold and plot the contemporaneous coefficient on policy uncertainty. The negative effect at the recall threshold ($-0.029$) steadily attenuates as $\tau$ rises, vanishing near the precision-optimized benchmark and reaching $0.001$ ($\text{SE} = 0.003$) at $\tau = 0.90$. Crucially, even as the number of admitted articles falls from roughly 851,000 to 58,000 across the grid (Panel C), the monthly regression sample is fixed and the confidence intervals in Panel A tighten, yielding increasingly precise null estimates rather than noisy attenuation.\footnote{This precision holds after accounting for the higher volatility of $\Delta \text{EPU}_t$ at stricter thresholds: scaling $\Delta \text{EPU}_t$ by its standard deviation at each threshold shifts the point estimate from $-0.0055$ ($\text{SE} = 0.0024$) at $\tau = 0.40$ to $0.0003$ ($\text{SE} = 0.0018$) at $\tau = 0.90$.}

Panel B explains this pattern by plotting the relative prevalence of four topic categories among the admitted articles, normalized to $\tau = 0.40$. Raising $\tau$ concentrates the index on routine financial and domestic economic coverage, more than doubling the relative frequency of stock market terms and raising general economic terms by 19 percent, while cutting military and foreign affairs terminology by 59 and 70 percent, respectively.\footnote{Panel B plots frequencies relative to $\tau = 0.40$, so categories start from different raw baselines (per 1,000 words at $\tau = 0.40$: general economic 22.48, foreign nations 8.41, military 5.73, and stock market 0.11, rising to 0.26 at $\tau = 0.90$). Thus, while stock market terms show the largest relative increase, the dominant volume shifts are the rise in general economic terms (to 26.84) and the declines in foreign nations and military terms (to 2.54 and 2.34).} The canonical negative relationship is therefore carried by articles near the boundary of the EPU definition, where economic policy intersects with war and geopolitical risk. Notably, to construct their historical index, BBD expanded their baseline policy dictionary to include terms such as \emph{war}, and \emph{tariff}---directly building this geopolitical and trade coverage into the benchmark series used by \citet{arouri}. 

\noindent By isolating only the most unambiguous domestic economic articles,  a stringent LLM threshold strips out the very shocks that correspond with the market response in the historical BBD index.

\section{Conclusion}\label{sec5}
Empirical research increasingly relies on predictive models to compress unstructured text into numerical regressors, yet the mapping from text to numbers is often treated as a fixed preprocessing step. Revisiting the BBD framework, we show that upstream choices across scoring models, decision thresholds, and aggregation rules shape not only measurement error $e^m_t$, but also the substantive composition $w^m_k$ of what an index admits. Fine-tuned LLMs recover nearly twice as many human-audited EPU articles as keyword rules, track human audit series far more closely out of sample, and overcome severe OCR noise to extend U.S. policy uncertainty measurement back to 1800. When we take these series to the stock-returns specification of \citet{arouri}, however, cleaner article-level classification does not simply reduce attenuation bias. Instead, moving from a keyword rule to a contextual model pools alarmist coverage with routine policy reporting, while tightening the threshold $\tau$ strips out war and geopolitical coverage in favor of domestic economic news, eliminating the negative contemporaneous return response.

These findings have immediate implications for the broader ecosystem of text-based uncertainty measures, which now spans dozens of country-specific indices and category-level sub-indices. Cross-country EPU comparisons rely on translated keyword dictionaries across languages with different idioms and journalistic conventions, while category sub-indices intersect additional term lists that compound false negatives and thin out article counts. Our results suggest that differences across countries or policy categories under dictionary rules may reflect shifts in the content boundary $B_{it}(\tau)$ rather than purely economic differences in the level or transmission of uncertainty. Modern contextual models, which transfer naturally across languages and can classify disaggregated policy categories within a single framework, offer a way to harmonize what an index admits across settings while keeping threshold and composition choices explicit.\footnote{The Supplemental Appendix explores the feasibility of both extensions. A single multi-label Longformer classifier trained on BBD's policy-category annotations exceeds BOW F1 by 40 to 132 percent in the four largest categories (Section IV). Applied to translations of the audit corpus into 29 languages, the fine-tuned Llama 3.1 (8B) classifier exceeds the F1 of the English BOW benchmark in every language; we use it to construct illustrative indices for Africa and Bangladesh from native-language news (Section VII).}

More broadly, our results highlight a gap in how applied economics treats machine-generated variables. A growing econometric literature develops valid inference for regressions using predicted variables, typically modeling prediction error as noise around a single target parameter. Yet when the underlying concept is latent and heterogeneous, changing the classifier or threshold changes the weighted average of treatment effects $\sum_k w^m_k \beta_k$ that a downstream regression identifies. Treating predictive pipelines as estimable parts of the econometric specification, by inspecting the composition of admitted observations and reporting sensitivity across thresholds and aggregation rules, is therefore essential for credible inference with unstructured data.
\clearpage

\bibliographystyle{aea}
\bibliography{refs}

@misc{battaglia2025,
      title={Inference for Regression with Variables Generated by AI or Machine Learning}, 
      author={Laura Battaglia and Timothy Christensen and Stephen Hansen and Szymon Sacher},
      year={2025},
      eprint={2402.15585},
      archivePrefix={arXiv},
      primaryClass={econ.EM},
      url={https://arxiv.org/abs/2402.15585}, 
}

@article{BBD_16,
   author = {Baker, Scott R. and Bloom, Nicholas and Davis, Steven J.},
   title = {MEASURING ECONOMIC POLICY UNCERTAINTY},
   journal = {The Quarterly journal of economics},
   volume = {131},
   number = {4},
   pages = {1593-1636},
                   ISSN = {0033-5533},
                   DOI = {10.1093/qje/qjw024},
                   year = {2016},
                   type = {Journal Article}
}

@article{shiller,
  title={Irrational exuberance: Revised and expanded third edition},
  author={Shiller, Robert J},
  year={2015},
  publisher={Princeton university press}
}

@article{arouri,
title = {Economic policy uncertainty and stock markets: Long-run evidence from the US},
journal = {Finance Research Letters},
volume = {18},
pages = {136-141},
year = {2016},
issn = {1544-6123},
doi = {https://doi.org/10.1016/j.frl.2016.04.011},
url = {https://www.sciencedirect.com/science/article/pii/S1544612316300435},
author = {Mohamed Arouri and Christophe Estay and Christophe Rault and David Roubaud}
}

@article{scikit-learn,
  title={Scikit-learn: Machine Learning in {P}ython},
  author={Pedregosa, F. and Varoquaux, G. and Gramfort, A. and Michel, V.
          and Thirion, B. and Grisel, O. and Blondel, M. and Prettenhofer, P.
          and Weiss, R. and Dubourg, V. and Vanderplas, J. and Passos, A. and
          Cournapeau, D. and Brucher, M. and Perrot, M. and Duchesnay, E.},
  journal={Journal of Machine Learning Research},
  volume={12},
  pages={2825--2830},
  year={2011}
}

@article{youden,
author = {Youden, W. J.},
title = {Index for rating diagnostic tests},
journal = {Cancer},
volume = {3},
number = {1},
pages = {32-35},
doi = {https://doi.org/10.1002/1097-0142(1950)3:1<32::AID-CNCR2820030106>3.0.CO;2-3},
url = {https://acsjournals.onlinelibrary.wiley.com/doi/abs/10.1002/1097-0142%281950%293%3A1%3C32%3A%3AAID-CNCR2820030106%3E3.0.CO%3B2-3},
eprint = {https://acsjournals.onlinelibrary.wiley.com/doi/pdf/10.1002/1097-0142%281950%293%3A1%3C32%3A%3AAID-CNCR2820030106%3E3.0.CO%3B2-3},
year = {1950}
}

@article{gentzkow_10,
   author = {Gentzkow, Matthew and Shapiro, Jesse M.},
   title = {What Drives Media Slant? Evidence From U.S. Daily Newspapers},
   journal = {Econometrica},
   volume = {78},
   number = {1},
   pages = {35-71},
   ISSN = {0012-9682},
   DOI = {10.3982/ECTA7195},
   year = {2010},
   type = {Journal Article}
}

@misc{dell_as,
   author = {Dell, Melissa and Carlson, Jacob and Bryan, Tom and Silcock, Emily and Arora, Abhishek and Shen, Zejiang and D'Amico-Wong, Luca and Le, Quan and Querubin, Pablo and Heldring, Leander},
   title = {American Stories: A Large-Scale Structured Text Dataset of Historical U.S. Newspapers},
   publisher = {Cornell University Library, arXiv.org},
   ISBN = {2331-8422},
   year = {2023},
   type = {Generic}
}

@article{dell_newswire,
   author = {Silcock, Emily and Arora, Abhishek and D'Amico-Wong, Luca and Dell, Melissa},
   title = {Newswire: A Large-Scale Structured Database of a Century of Historical News},
   DOI = {10.48550/arxiv.2406.09490},
   year = {2024},
   type = {Journal Article}
}

@TechReport{baker_2019,
type={NBER Working Papers},
institution={National Bureau of Economic Research, Inc},
author={Scott R. Baker and Nicholas Bloom and Steven J. Davis and Kyle J. Kost},
title={Policy News and Stock Market Volatility},
year={2019},
month={Mar},
number={25720},
doi={None},
url={https://ideas.repec.org/p/nbr/nberwo/25720.html},
}

@article{stone,
  title={The general inquirer: A computer approach to content analysis.},
  author={Philip J. Stone and Dexter C. Dunphy and Marshall S. Smith},
  journal={American Educational Research Journal},
  year={1967},
  volume={4},
  pages={397},
  url={https://api.semanticscholar.org/CorpusID:60936250}
}

@article{newey,
 ISSN = {00129682, 14680262},
 URL = {http://www.jstor.org/stable/1913610},
 author = {Whitney K. Newey and Kenneth D. West},
 journal = {Econometrica},
 number = {3},
 pages = {703--708},
 publisher = {[Wiley, Econometric Society]},
 title = {A Simple, Positive Semi-Definite, Heteroskedasticity and Autocorrelation Consistent Covariance Matrix},
 urldate = {2026-10-23},
 volume = {55},
 year = {1987}
}

@misc{araci,
      title={FinBERT: Financial Sentiment Analysis with Pre-trained Language Models}, 
      author={Dogu Araci},
      year={2019},
      eprint={1908.10063},
      archivePrefix={arXiv},
      primaryClass={cs.CL},
      url={https://arxiv.org/abs/1908.10063}, 
}

@article{optuna,
   author = {Akiba, Takuya and Sano, Shotaro and Yanase, Toshihiko and Ohta, Takeru and Koyama, Masanori},
   title = {Optuna: A Next-generation Hyperparameter Optimization Framework},
   journal = {arXiv.org},
   ISSN = {2331-8422},
   DOI = {10.48550/arxiv.1907.10902},
   year = {2019},
   type = {Journal Article}
}

@article{vafa2024career,
   author = {Vafa, Keyon and Palikot, Emil and Du, Tianyu and Kanodia, Ayush and Athey, Susan and Blei, David M.},
   title = {CAREER: A Foundation Model for Labor Sequence Data},
   DOI = {10.48550/arxiv.2202.08370},
   year = {2022},
   type = {Journal Article}
}

@article{dulabor,
   author = {Athey, Susan and Brunborg, Herman and Du, Tianyu and Kanodia, Ayush and Vafa, Keyon},
   title = {LABOR-LLM: Language-Based Occupational Representations with Large Language Models},
   DOI = {10.48550/arxiv.2406.17972},
   year = {2024},
   type = {Journal Article}
}

@article{charemza_bow,
   author = {Charemza, Wojciech and Makarova, Svetlana and Rybiński, Krzysztof},
   title = {Economic uncertainty and natural language processing; The case of Russia},
   journal = {Economic analysis and policy},
   volume = {73},
   pages = {546-562},
   ISSN = {0313-5926},
   DOI = {10.1016/j.eap.2021.11.011},
   year = {2022},
   type = {Journal Article}
}

@article{choi,
   author = {Benguria, Felipe and Choi, Jaerim and Swenson, Deborah L. and Xu, Mingzhi},
   title = {Anxiety or pain? The impact of tariffs and uncertainty on Chinese firms in the trade war},
   journal = {Journal of international economics},
   volume = {137},
   pages = {103608},
   note = {},
   ISSN = {0022-1996},
   DOI = {10.1016/j.jinteco.2022.103608},
   year = {2022},
   type = {Journal Article}
}

@article{azqueta,
   author = {Azqueta-Gavaldón, Andrés and Hirschbühl, Dominik and Onorante, Luca and Saiz, Lorena},
   title = {Sources of Economic Policy Uncertainty in the euro area},
   journal = {European economic review},
   volume = {152},
   pages = {104373},
   ISSN = {0014-2921},
   DOI = {10.1016/j.euroecorev.2023.104373},
   year = {2023},
   type = {Journal Article}
}

@article{tobback,
   author = {Tobback, Ellen and Naudts, Hans and Daelemans, Walter and Junqué de Fortuny, Enric and Martens, David},
   title = {Belgian economic policy uncertainty index: Improvement through text mining},
   journal = {International journal of forecasting},
   volume = {34},
   number = {2},
   pages = {355-365},
   ISSN = {0169-2070},
   DOI = {10.1016/j.ijforecast.2016.08.006},
   year = {2018},
   type = {Journal Article}
}

@techreport{ludwig,
 title = "Large Language Models: An Applied Econometric Framework",
 author = "Ludwig, Jens and Mullainathan, Sendhil and Rambachan, Ashesh",
 institution = "National Bureau of Economic Research",
 type = "Working Paper",
 series = "Working Paper Series",
 number = "33344",
 year = "2025",
 month = "January",
 doi = {10.3386/w33344},
 URL = "http://www.nber.org/papers/w33344",
}

@techreport{audrino,
  author       = {Audrino, Francesco and Maly, Jessica and Stalder, Simon},
  title        = {Quantifying Uncertainty: A New Era of Measurement through Large Language Models},
  journal      = {Swiss Finance Institute Research Paper},
  year         = {2024},
  number       = {24-68},
  month        = {July},
  doi          = {10.2139/ssrn.4903414},
  url          = {https://ssrn.com/abstract=4903414}
}

@article{ito,
title = {A novel content-based approach to measuring monetary policy uncertainty using fine-tuned LLMs},
journal = {Finance Research Letters},
volume = {75},
pages = {106832},
year = {2025},
issn = {1544-6123},
doi = {https://doi.org/10.1016/j.frl.2025.106832},
url = {https://www.sciencedirect.com/science/article/pii/S1544612325000972},
author = {Arata Ito and Masahiro Sato and Rui Ota}
}

@article{trust,
   author = {Trust, Paul and Zahran, Ahmed and Minghim, Rosane},
   title = {Understanding the influence of news on society decision making: application to economic policy uncertainty},
   journal = {Neural computing \& applications},
   volume = {35},
   number = {20},
   pages = {14929-14945},
   ISSN = {0941-0643},
   DOI = {10.1007/s00521-023-08438-8},
   year = {2023},
   type = {Journal Article}
}

@article{kong,
   author = {Kong, Qunxi and Li, Rongrong and Wang, Ziqi and Peng, Dan},
   title = {Economic policy uncertainty and firm investment decisions: Dilemma or opportunity?},
   journal = {International review of financial analysis},
   volume = {83},
   pages = {102301},
   ISSN = {1057-5219},
   DOI = {10.1016/j.irfa.2022.102301},
   year = {2022},
   type = {Journal Article}
}

@article{kitsul,
   author = {Jahan-Parvar, Mohammad R. and Kitsul, Yuriy and Rahman, Jamil and Wilson, Beth Anne},
   title = {Foreign economic policy uncertainty and U.S. equity returns},
   journal = {International finance discussion papers},
   number = {1401},
   pages = {1-40},
   ISSN = {1073-2500},
   DOI = {10.17016/ifdp.2024.1401},
   year = {2024},
   type = {Journal Article}
}

@article{caldara,
   author = {Caldara, Dario and Iacoviello, Matteo},
   title = {Measuring Geopolitical Risk},
   journal = {The American economic review},
   volume = {112},
   number = {4},
   pages = {1194-1225},
   ISSN = {0002-8282},
   DOI = {10.1257/aer.20191823},
   year = {2022},
   type = {Journal Article}
}

@article{thaqeb,
   author = {Al-Thaqeb, Saud Asaad and Algharabali, Barrak Ghanim},
   title = {Economic policy uncertainty: A literature review},
   journal = {Journal of economic asymmetries},
   volume = {20},
   pages = {e00133},
   ISSN = {1703-4949},
   DOI = {10.1016/j.jeca.2019.e00133},
   year = {2019},
   type = {Journal Article}
}

@article{gentzkow_tad,
   author = {Gentzkow, Matthew and Kelly, Bryan and Taddy, Matt},
   title = {Text as Data},
   journal = {Journal of economic literature},
   volume = {57},
   number = {3},
   pages = {535-574},
   ISSN = {0022-0515},
   DOI = {10.1257/jel.20181020},
   year = {2019},
   type = {Journal Article}
}

@article{keith,
   author = {Keith, Katherine A. and Teichmann, Christoph and O'Connor, Brendan and Meij, Edgar},
   title = {Uncertainty over Uncertainty: Investigating the Assumptions, Annotations, and Text Measurements of Economic Policy Uncertainty},
   journal = {arXiv.org},
   ISSN = {2331-8422},
   DOI = {10.48550/arxiv.2010.04706},
   year = {2020},
   type = {Journal Article}
}

@misc{noailly,
   author = {Noailly, Joelle and Nowzohour, Laura M. and van den Heuvel, Matthias},
   title = {Does Environmental Policy Uncertainty Hinder Investments Towards a Low-Carbon Economy?},
   publisher = {National Bureau of Economic Research},
   ISBN = {0898-2937},
   DOI = {10.3386/w30361},
   year = {2022},
   type = {Generic}
}

@inproceedings{vaswani,
 author = {Vaswani, Ashish and Shazeer, Noam and Parmar, Niki and Uszkoreit, Jakob and Jones, Llion and Gomez, Aidan N and Kaiser, \L ukasz and Polosukhin, Illia},
 booktitle = {Advances in Neural Information Processing Systems},
 editor = {I. Guyon and U. Von Luxburg and S. Bengio and H. Wallach and R. Fergus and S. Vishwanathan and R. Garnett},
 pages = {},
 publisher = {Curran Associates, Inc.},
 title = {Attention is All you Need},
 url = {https://proceedings.neurips.cc/paper_files/paper/2017/file/3f5ee243547dee91fbd053c1c4a845aa-Paper.pdf},
 volume = {30},
 year = {2017}
}

@article{devlin,
   author = {Devlin, Jacob and Ming-Wei, Chang and Lee, Kenton and Toutanova, Kristina},
   title = {BERT: Pre-training of Deep Bidirectional Transformers for Language Understanding},
   journal = {arXiv.org},
   ISSN = {2331-8422},
   DOI = {10.48550/arxiv.1810.04805},
   year = {2019},
   type = {Journal Article}
}

@article{brown,
   author = {Brown, Tom B. and Mann, Benjamin and Ryder, Nick and Subbiah, Melanie and Kaplan, Jared and Dhariwal, Prafulla and Neelakantan, Arvind and Shyam, Pranav and Sastry, Girish and Askell, Amanda and Agarwal, Sandhini and Herbert-Voss, Ariel and Krueger, Gretchen and Henighan, Tom and Child, Rewon and Ramesh, Aditya and Ziegler, Daniel M. and Wu, Jeffrey and Winter, Clemens and Hesse, Christopher and Chen, Mark and Sigler, Eric and Litwin, Mateusz and Gray, Scott and Chess, Benjamin and Clark, Jack and Berner, Christopher and McCandlish, Sam and Radford, Alec and Sutskever, Ilya and Amodei, Dario},
   title = {Language Models are Few-Shot Learners},
   DOI = {10.48550/arxiv.2005.14165},
   year = {2020},
   type = {Journal Article}
}

@article{gilardi,
   author = {Gilardi, Fabrizio and Alizadeh, Meysam and Kubli, Maël},
   title = {ChatGPT outperforms crowd workers for text-annotation tasks},
   journal = {Proceedings of the National Academy of Sciences - PNAS},
   volume = {120},
   number = {30},
   pages = {e2305016120},
   ISSN = {0027-8424},
   DOI = {10.1073/pnas.2305016120},
   year = {2023},
   type = {Journal Article}
}

@article{touvron,
   author = {Touvron, Hugo and Lavril, Thibaut and Izacard, Gautier and Martinet, Xavier and Lachaux, Marie-Anne and Lacroix, Timothée and Rozière, Baptiste and Goyal, Naman and Hambro, Eric and Azhar, Faisal and Rodriguez, Aurelien and Joulin, Armand and Grave, Edouard and Lample, Guillaume},
   title = {LLaMA: Open and Efficient Foundation Language Models},
   DOI = {10.48550/arxiv.2302.13971},
   year = {2023},
   type = {Journal Article}
}

@misc{lin,
   author = {Lin, Stephanie and Hilton, Jacob and Evans, Owain},
   title = {TruthfulQA: Measuring How Models Mimic Human Falsehoods},
   publisher = {Cornell University Library, arXiv.org},
   ISBN = {2331-8422},
   year = {2022},
   type = {Generic}
}

@misc{zhao,
      title={Calibrate Before Use: Improving Few-Shot Performance of Language Models}, 
      author={Tony Z. Zhao and Eric Wallace and Shi Feng and Dan Klein and Sameer Singh},
      year={2021},
      eprint={2102.09690},
      archivePrefix={arXiv},
      primaryClass={cs.CL},
      url={https://arxiv.org/abs/2102.09690}, 
}

@misc{ullman,
   author = {Ullman, Tomer},
   title = {Large Language Models Fail on Trivial Alterations to Theory-of-Mind Tasks},
   publisher = {Cornell University Library, arXiv.org},
   ISBN = {2331-8422},
   year = {2023},
   type = {Generic}
}

@article{beltagy,
   author = {Beltagy, Iz and Peters, Matthew E. and Cohan, Arman},
   title = {Longformer: The Long-Document Transformer},
   journal = {arXiv.org},
   ISSN = {2331-8422},
   DOI = {10.48550/arxiv.2004.05150},
   year = {2020},
   type = {Journal Article}
}

@misc{llama3,
      title={The Llama 3 Herd of Models}, 
      author={Llama Team, AI @ Meta},
      year={2024},
      eprint={2407.21783},
      archivePrefix={arXiv},
      primaryClass={cs.AI},
      url={https://arxiv.org/abs/2407.21783}, 
}

@article{epping2025harnessing,
author = {Epping, Gunnar P. and Caplin, Andrew and Duhaime, Erik and Holmes, William R. and Martin, Daniel and Trueblood, Jennifer S.},
language = {eng},
issn = {1545-8490},
journal = {Decision analysis},
title = {Harnessing Human Uncertainty to Train More Accurate and Aligned AI Systems},
year = {2026},
}

@misc{shap,
   author = {Lundberg, Scott M. and Lee, Su-In},
   title = {A unified approach to interpreting model predictions},
   publisher = {Curran Associates Inc.},
   pages = {4768–4777},
   year = {2017},
   type = {Conference Paper}
}

@article{cunningham,
   author = {Cunningham, Hoagy and Ewart, Aidan and Riggs, Logan and Huben, Robert and Sharkey, Lee},
   title = {Sparse autoencoders find highly interpretable features in language models},
   journal = {arXiv preprint arXiv:2309.08600},
   year = {2023},
   type = {Journal Article}
}

@misc{dunefsky,
   author = {Dunefsky, Jacob and Chlenski, Philippe and Nanda, Neel},
   title = {Transcoders find interpretable LLM feature circuits},
   publisher = {Curran Associates Inc.},
   volume = {37},
   pages = {Article 768},
   year = {2024},
   type = {Conference Paper}
}

@article{tetlock,
issn = {0022-1082},
author = {Tetlock, Paul C.},
address = {Malden, USA},
copyright = {Copyright 2007 The American Finance Association},
journal = {The Journal of finance (New York)},
language = {eng},
number = {3},
pages = {1139-1168},
publisher = {Blackwell Publishing Inc},
title = {Giving Content to Investor Sentiment: The Role of Media in the Stock Market},
volume = {62},
year = {2007},
}

@article{ahir,
   author = {Ahir, Hites and Bloom, Nicholas and Furceri, Davide},
   title = {The World Uncertainty Index},
   journal = {National Bureau of Economic Research Working Paper Series},
   volume = {No. 29763},
   DOI = {10.3386/w29763},
   url = {http://www.nber.org/papers/w29763},
   year = {2022},
   type = {Journal Article}
}

@misc{carlson,
      title={A Unifying Framework for Robust and Efficient Inference with Unstructured Data}, 
      author={Jacob Carlson and Melissa Dell},
      year={2025},
      eprint={2505.00282},
      archivePrefix={arXiv},
      primaryClass={econ.EM},
      url={https://arxiv.org/abs/2505.00282}, 
}

@TechReport{hong,
type={IMF Working Papers},
institution={International Monetary Fund},
author={Mr. Gee Hee Hong and Shikun (Barry) Ke and Anh D. M. Nguyen},
title={The Economic Impact of Fiscal Policy Uncertainty: Evidence from a New Cross-Country Database},
year={2024},
month={Sep},
number={2024/209},
doi={None},
url={https://ideas.repec.org/p/imf/imfwpa/2024-209.html},
}

@article{diakonova,
   author = {Diakonova, Marina and Ghirelli, Corinna and Wu, Juan Quiñónez},
   title = {Economic Policy Uncertainty in Central America and the Dominican Republic},
   journal = {Latin American journal of central banking},
   pages = {100166},
   ISSN = {2666-1438},
   DOI = {10.1016/j.latcb.2025.100166},
   year = {2025},
   type = {Journal Article}
}

@article{loughran,
   author = {Loughran, Tim and McDonald, Bill},
   title = {When Is a Liability Not a Liability? Textual Analysis, Dictionaries, and 10-Ks},
   journal = {The Journal of finance (New York)},
   volume = {66},
   number = {1},
   pages = {35-65},
   ISSN = {0022-1082},
   DOI = {10.1111/j.1540-6261.2010.01625.x},
   year = {2011},
   type = {Journal Article}
}

\clearpage

\setcounter{section}{0}
\setcounter{table}{0}
\setcounter{figure}{0}
\setcounter{equation}{0}
\renewcommand{\thetable}{S\arabic{table}}
\renewcommand{\thefigure}{S\arabic{figure}}
\renewcommand{\theequation}{S\arabic{equation}}

\begin{center}
    {\Large \bfseries Supplemental Appendix to ``Narratives to Numbers: Large Language Models and Economic Policy Uncertainty''}
\end{center}
\vspace{1em}

\begin{bibunit}[aea]

\section{Audit Data and Sample Splits}\label{sec:data}
The BBD audit covers approximately 12{,}000 newspaper articles from 1900--2012, each labeled by human auditors as EPU-related or not and supplemented by auditor certainty scores and policy category annotations. We remove duplicate entries and articles without recoverable text, leaving 10{,}393 labeled articles. Table~\ref{tab:data_split} reports the composition of each sample split. The random split assigns articles to training (70 percent), validation (20 percent), and test (10 percent) sets at random. The temporal split assigns articles from 1900--2005 to training and validation and holds out all articles from 2006--2012 as the test set, which provides the contiguous out-of-sample window used to evaluate index fidelity in Section III.A of the main text. Title-only models use the same partitions; counts are slightly lower because some articles lack a recoverable title.

\begin{table}[!htbp]
    \caption{Data Splits for EPU Classification}
    \centering
    \resizebox{0.8\textwidth}{!}{%
    \begin{threeparttable}
    \begin{tabular}{lcccccc}
    \toprule
    \multicolumn{7}{c}{\textit{Full articles}} \\
    \midrule
    & \multicolumn{3}{c}{Random} & \multicolumn{3}{c}{Temporal} \\
    \cmidrule(lr){2-4} \cmidrule(lr){5-7}
    & Articles & EPU = 0 & EPU = 1 & Articles & EPU = 0 & EPU = 1 \\
    \midrule
    Train       & 7{,}275 & 4{,}081 & 3{,}194 & 7{,}660 & 4{,}279 & 3{,}381 \\
    Validation  & 2{,}078 & 1{,}159 &   919   & 1{,}915 & 1{,}012 &   903   \\
    Test        & 1{,}040 &   546   &   494   &   818   &   495   &   323   \\
    \midrule
    \multicolumn{7}{c}{\textit{Titles}} \\
    \midrule
    & \multicolumn{3}{c}{Random} & \multicolumn{3}{c}{Temporal} \\
    \cmidrule(lr){2-4} \cmidrule(lr){5-7}
    & Articles & EPU = 0 & EPU = 1 & Articles & EPU = 0 & EPU = 1 \\
    \midrule
    Train       & 7{,}230 & 3{,}988 & 3{,}242 & 7{,}612 & 4{,}182 & 3{,}430 \\
    Validation  & 2{,}065 & 1{,}154 &   911   & 1{,}904 & 1{,}037 &   867   \\
    Test        & 1{,}034 &   570   &   464   &   813   &   493   &   320   \\
    \bottomrule
    \end{tabular}
    \begin{tablenotes}[flushleft]
        \footnotesize \item Notes: The random split assigns articles to training, validation, and test sets at random. The temporal split assigns articles from 1900--2005 to training and validation and articles from 2006--2012 to the test set. EPU = 1 indicates articles that auditors labeled EPU-related.
    \end{tablenotes}
    \end{threeparttable}
    }%
    \label{tab:data_split}
\end{table}

\clearpage

\section{Model Training}\label{sec:training}
We select hyperparameters with Optuna's Tree-structured Parzen Estimator (TPE) sampler \citep{optuna}. The TPE sampler fits separate density estimates to hyperparameter configurations associated with relatively low validation loss, $f(x)$, and to all remaining configurations, $g(x)$, and draws each new trial to maximize the ratio $f(x)/g(x)$. Within each trial, the model is trained to minimize the binary cross-entropy loss
\begin{align}
    \mathrm{BCE}
    &= - \frac{1}{N} \sum_{i=1}^N \Big( y_i \log(p_i) + \big(1 - y_i\big) \log\big(1-p_i\big) \Big), \label{eq:bce}
\end{align}
where $y_i$ is the auditor label of article $i$ and $p_i$ is the predicted probability that $y_i = 1$. Table~\ref{tab:hyperparams} lists the search space. For each model and context length, we train at least 15 trials and select the trial with the lowest validation loss, for a minimum of roughly 576 GPU training hours across all specifications. Training uses the Hugging Face \texttt{Trainer} API \citep{huggingface}. We release the selected models on the Hugging Face Hub, and Section~\ref{sec:code} shows example code. Scripts for data cleaning, splitting, training, and evaluation will be made available in a public repository at the time of publication.

\begin{table}[!htbp]
    \caption{Hyperparameter Search Space}
    \centering
    \begin{threeparttable}
    \begin{tabular}{lcl}
    \toprule
    Hyperparameter & Range & Models \\
    \midrule
    Learning rate        & [$10^{-8}$, $10^{-4}$] & All \\
    Weight decay         & [$10^{-5}$, 0.9]       & All \\
    Gradient clipping    & [0, 1]                 & All \\
    Warmup ratio         & [0, 0.1]               & All \\
    Dropout probability  & [0, 0.9]               & All \\
    \addlinespace
    LoRA rank            & [8, 128]               & Llama 3.1 (8B) \\
    LoRA scaling factor $\alpha$ & [8, 256]       & Llama 3.1 (8B) \\
    LoRA dropout         & [0, 0.9]               & Llama 3.1 (8B) \\
    \bottomrule
    \end{tabular}
    \begin{tablenotes}[flushleft]
        \footnotesize \item Notes: Ranges searched by the Optuna TPE sampler for every model and context-length combination. LoRA parameters apply only to the parameter-efficient fine-tuning of Llama 3.1 (8B) described in Section~\ref{sec:peft}.
    \end{tablenotes}
    \end{threeparttable}
    \label{tab:hyperparams}
\end{table}

We implement the SVM benchmarks in scikit-learn \citep{scikit-learn} and evaluate every combination of regularization strength [0.1, 1, 10, 100, 500, 1000], kernel [linear, RBF, polynomial, sigmoid], kernel coefficient $\gamma$ [scale, auto], and text representation [TF-IDF, raw counts].

\subsection{Parameter-Efficient Fine-Tuning}\label{sec:peft}

Larger models such as Llama 3.1 (8B) require substantially more computational resources to train. To limit this overhead, we use Low-Rank Adaptation (LoRA), which applies a low-rank decomposition to selected weight matrices and learns small ``adapter'' updates while keeping the original weights frozen \citep{lora}.\footnote{We incorporate the rank-stabilized scaling factor of \citet{rslora}, which mitigates performance degradation for higher-rank adapters. Additional efficiency gains are possible through 4-bit quantization \citep{qlora}, which we do not use.} The LoRA implementation in the Hugging Face PEFT library introduces three additional hyperparameters, the adapter rank, the scaling factor $\alpha$, and the adapter dropout, which we tune over the ranges in Table~\ref{tab:hyperparams} \citep{peft}. Many strategies for combining parameter-efficient fine-tuning and quantization are possible; we adopt a simple and accessible configuration rather than exhausting the design space. Figure~\ref{fig:llm_diagram} summarizes the pipeline.

\begin{figure}[!htbp]
    \centering
    \includegraphics[width=\textwidth]{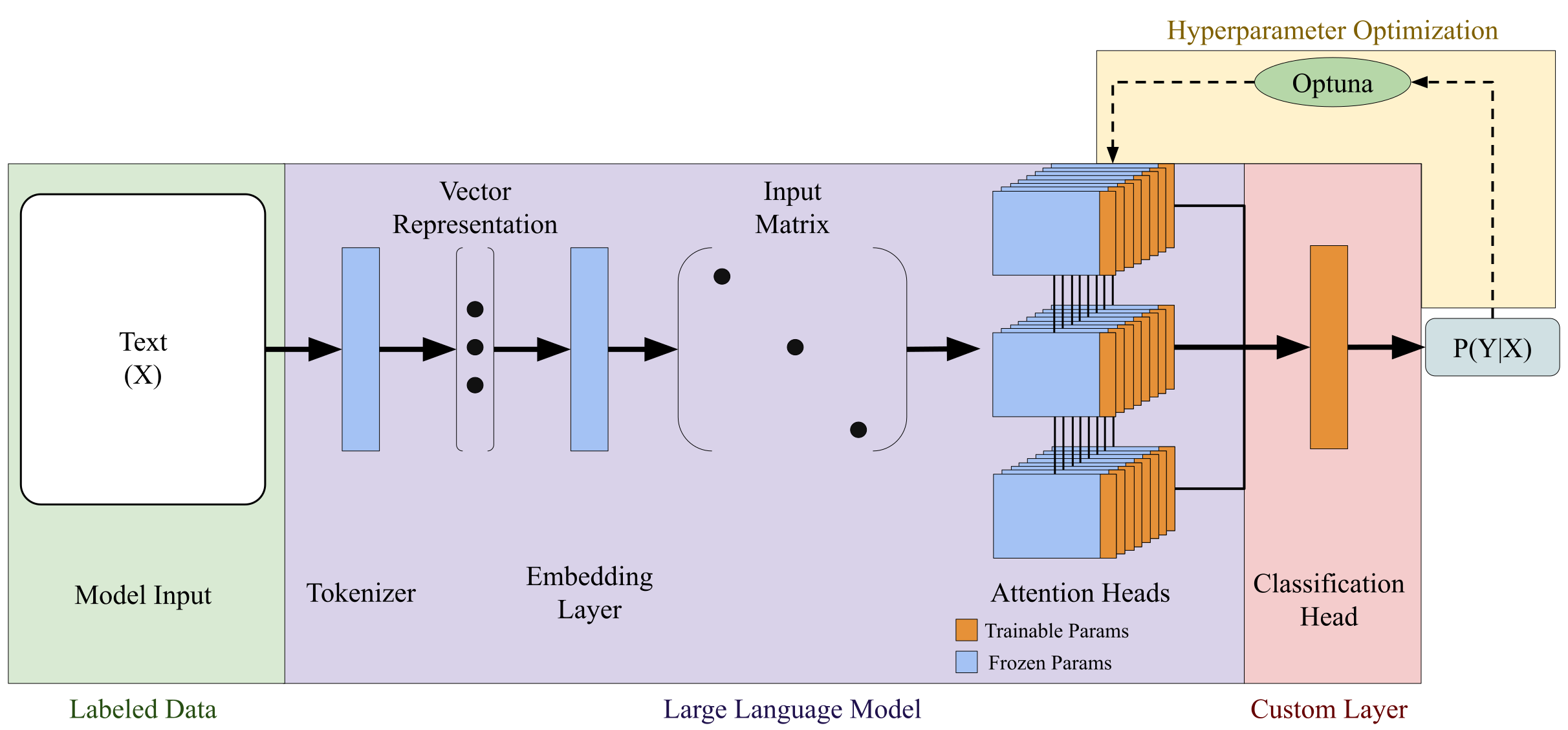}
    \caption{LLM Fine-Tuning Pipeline for EPU Classification}
    \label{fig:llm_diagram}
    \fignote{The diagram traces the pipeline from raw text to parameter-efficient fine-tuning. The final model returns article-level probabilities $\hat{p}_{ijt}$ that can be thresholded or aggregated into an index (Definition 1 in the main text).}
\end{figure}

\clearpage

\section{Additional Classifier Results}\label{sec:classifiers}
Table~\ref{tab:perf_by_split} reports test-set performance for every architecture and context length under both sample splits, with decision thresholds selected by Youden's index on each split's validation set. Performance rises with context length for Longformer and Llama 3.1 (8B). BERT is competitive at short inputs but cannot extend beyond 512 tokens, so its best specification falls short of the 2048-token Longformer and Llama 3.1 (8B) models in F1 and AUROC under both splits.

Figure~\ref{fig:split_comparison} compares the two splits directly. Models trained only on articles through 2005 perform comparably on the 2006--2012 test set to models trained and tested on random draws from the full period. At 2048 tokens, Longformer reaches an F1 of 77.07 under the temporal split against 76.65 under the random split, and Llama 3.1 (8B) reaches 77.16 against 77.52. AUROC is higher under the temporal split at most context lengths. Because the two test sets contain different articles, these comparisons reflect test-set composition as well as the split itself; however, neither suggests that performance deteriorates when a classifier trained on earlier text is applied to later coverage. Figure~\ref{fig:roc_all} plots the corresponding ROC curves.

\begin{table}[!htbp]
    \caption{Classifier Performance by Context Length and Sample Split}
    \centering
    \resizebox{\textwidth}{!}{%
    \begin{threeparttable}
    \begin{tabular}{l*{10}{c}}
\toprule
 & \multicolumn{5}{c}{Random split} & \multicolumn{5}{c}{Temporal split} \\
\cmidrule(lr){2-6}\cmidrule(lr){7-11}
Context length & Acc. & Prec. & Rec. & F1 & AUROC & Acc. & Prec. & Rec. & F1 & AUROC \\
\midrule
\multicolumn{11}{l}{\textit{BERT}} \\
32 (Title) & 61.03 & 55.62 & 65.09 & 59.98 & 64.94 & 68.88 & 61.59 & 55.62 & 58.46 & 73.79 \\
128 & 67.98 & 65.94 & 67.41 & 66.67 & 75.29 & 73.84 & 64.46 & 75.23 & 69.43 & 80.69 \\
256 & 73.85 & 70.48 & 77.33 & 73.75 & 78.86 & 72.49 & 61.19 & 82.97 & 70.43 & 83.92 \\
512 & 73.75 & 70.35 & 77.33 & 73.67 & 81.07 & 75.43 & 66.31 & 76.78 & 71.16 & 84.37 \\
\addlinespace
\multicolumn{11}{l}{\textit{Longformer}} \\
32 (Title) & 59.67 & 54.62 & 59.91 & 57.14 & 64.62 & 69.13 & 60.06 & 64.38 & 62.14 & 76.34 \\
128 & 70.29 & 69.39 & 67.00 & 68.18 & 76.66 & 73.35 & 64.71 & 71.52 & 67.94 & 80.98 \\
256 & 70.38 & 67.35 & 73.08 & 70.10 & 78.06 & 74.21 & 64.00 & 79.26 & 70.82 & 83.99 \\
512 & 73.65 & 71.07 & 75.10 & 73.03 & 81.61 & 77.75 & 70.09 & 76.16 & 73.00 & 86.38 \\
1024 & 74.04 & 69.11 & 81.98 & 75.00 & 83.70 & 78.97 & 71.39 & 78.02 & 74.56 & 88.81 \\
2048 & 76.15 & 71.65 & 82.39 & 76.65 & 84.96 & 80.07 & 70.62 & 84.83 & 77.07 & 89.80 \\
\addlinespace
\multicolumn{11}{l}{\textit{Llama 3.1 (8B)}} \\
32 (Title) & 56.67 & 51.44 & 61.64 & 56.08 & 60.10 & 68.39 & 59.24 & 63.12 & 61.12 & 72.59 \\
128 & 68.75 & 66.41 & 69.23 & 67.79 & 76.18 & 74.08 & 65.37 & 73.07 & 69.01 & 80.09 \\
256 & 72.50 & 69.48 & 75.10 & 72.18 & 80.17 & 74.94 & 66.39 & 73.99 & 69.99 & 82.47 \\
512 & 77.12 & 74.43 & 78.95 & 76.62 & 83.99 & 76.53 & 67.95 & 76.78 & 72.09 & 82.68 \\
1024 & 77.12 & 74.06 & 79.76 & 76.80 & 84.43 & 77.87 & 69.83 & 77.40 & 73.42 & 83.39 \\
2048 & 78.37 & 76.53 & 78.54 & 77.52 & 86.24 & 81.54 & 75.44 & 78.95 & 77.16 & 90.68 \\
\bottomrule
\end{tabular}

    \begin{tablenotes}[flushleft]
        \footnotesize \item Notes: Test-set accuracy, precision, recall, F1, and AUROC, in percent, at Youden-optimized thresholds selected on each split's validation set. The random-split test set contains 1{,}040 articles; the temporal-split test set contains the 818 articles from 2006--2012. Title models use article titles only. BERT is limited to 512 tokens.
    \end{tablenotes}
    \end{threeparttable}
    }%
    \label{tab:perf_by_split}
\end{table}

\begin{figure}[!htbp]
    \centering
    \includegraphics[width=\textwidth]{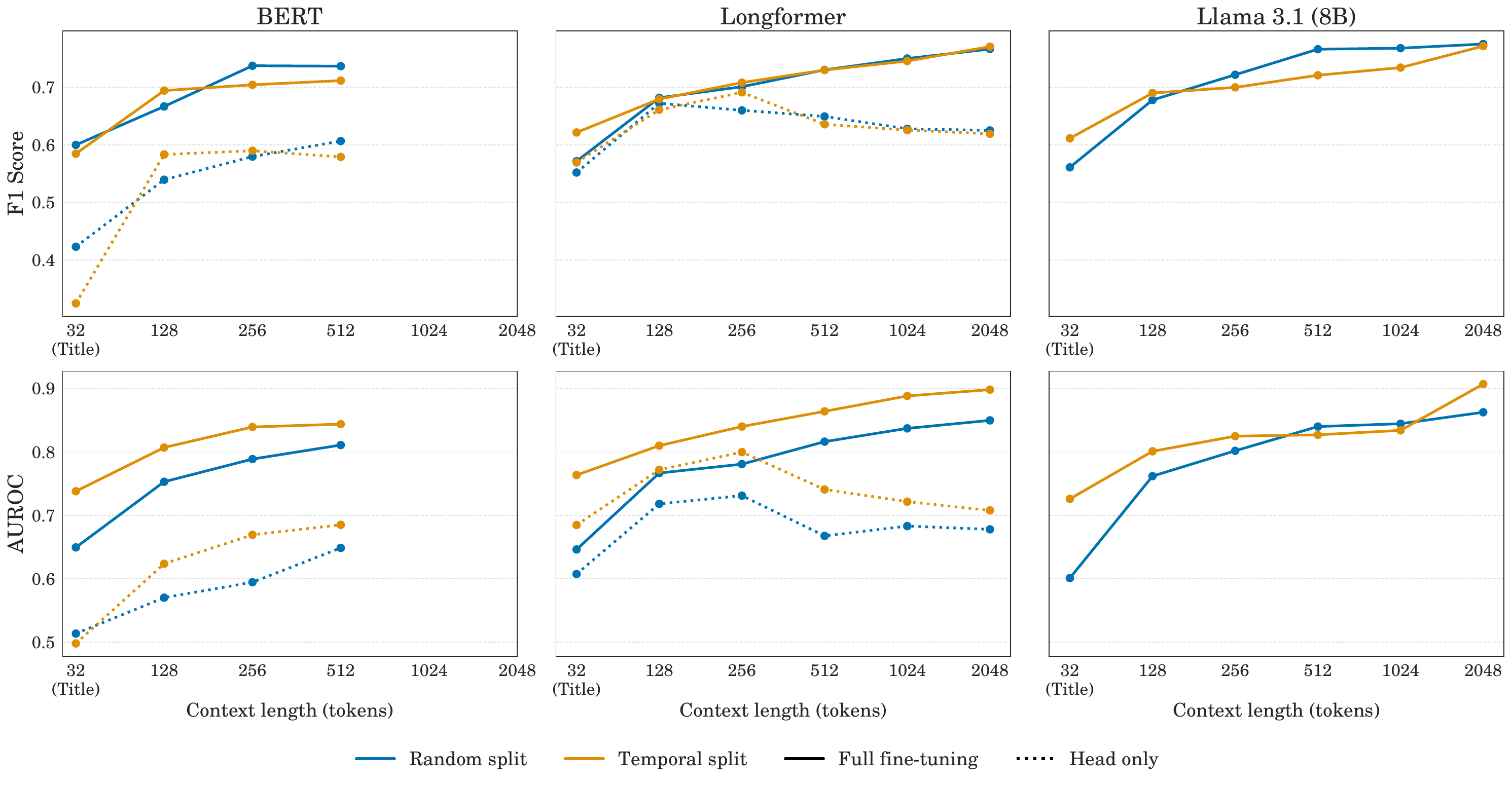}
    \caption{Random and Temporal Splits by Context Length}
    \label{fig:split_comparison}
    \fignote{Test-set F1 (top row) and AUROC (bottom row) at Youden-optimized thresholds. Blue lines use the random split and orange lines the temporal split. Solid lines fine-tune the backbone and classification head; dotted lines train only the classification head (BERT and Longformer). The random-split test set contains 1{,}040 articles drawn from 1900--2012; the temporal-split test set contains 818 articles from 2006--2012.}
\end{figure}

\begin{figure}[!htbp]
    \centering
    \includegraphics[width=0.85\textwidth]{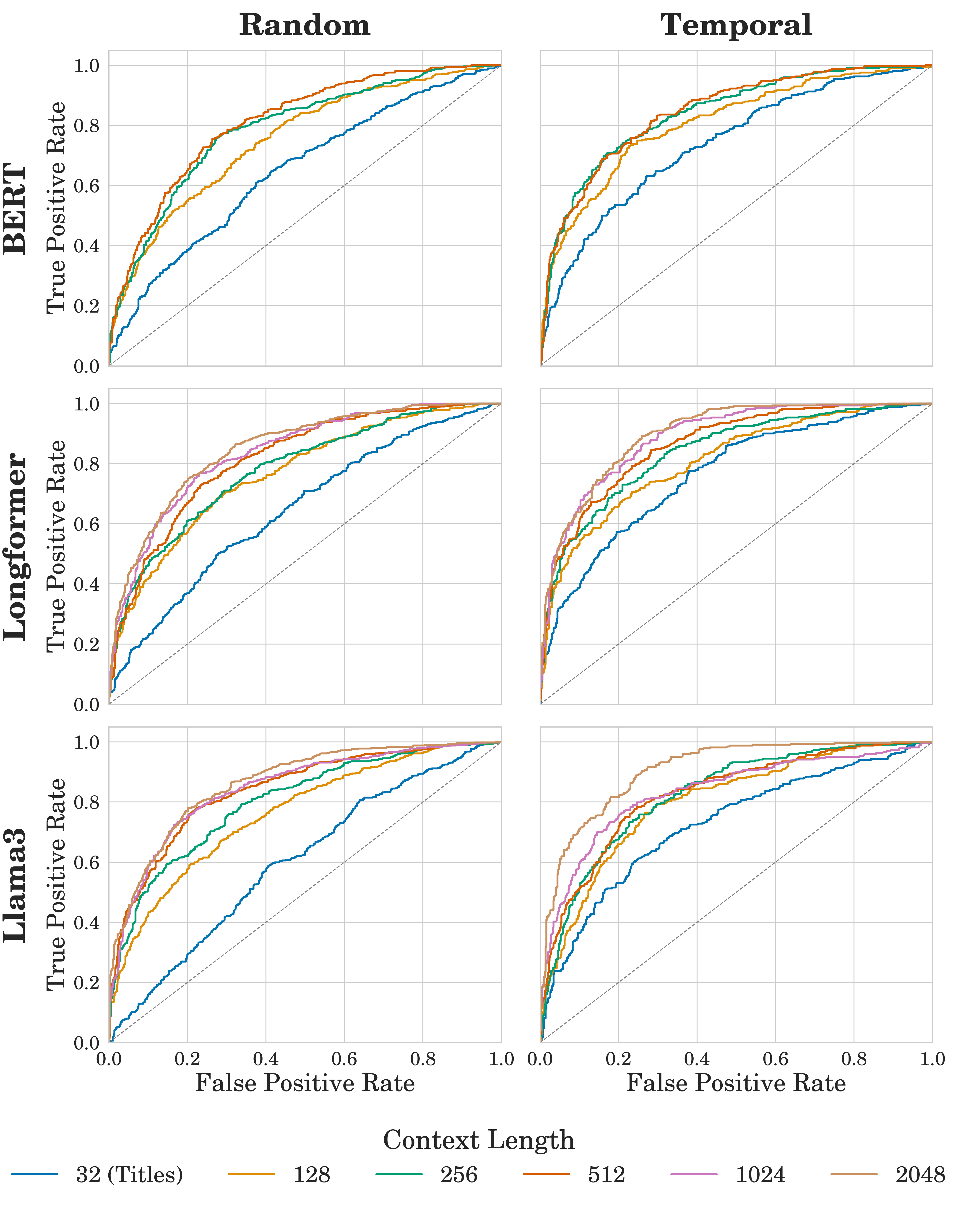}
    \caption{ROC Curves by Model, Context Length, and Sample Split}
    \label{fig:roc_all}
    \fignote{Each panel plots the true positive rate against the false positive rate on the held-out test set, with one curve per context length. Rows correspond to BERT, Longformer, and Llama 3.1 (8B) (labeled Llama3); columns correspond to the random and temporal splits.}
\end{figure}

\clearpage

\subsection{Fine-Tuning the Backbone}

In the main specifications we fine-tune both the LLM backbone and the classification head. Freezing the backbone and training only the head amounts to estimating a logit on fixed pre-trained representations $\tilde{C}$. Table~\ref{tab:head_vs_full} and Figure~\ref{fig:roc_head} compare the two. Head-only training lowers F1 and AUROC at every context length for both BERT and Longformer, and for Longformer the gap widens with input length. For Longformer (2048) on the random split, F1 falls from 76.65 to 62.50 and AUROC from 84.96 to 67.79. Head-only performance peaks at 128 to 256 tokens and declines thereafter, so additional text improves classification only when the backbone can adapt to the cues most informative about EPU.

\begin{table}[!htbp]
    \caption{Full Fine-Tuning versus Classification Head Only}
    \centering
    \resizebox{\textwidth}{!}{%
    \begin{threeparttable}
    \begin{tabular}{l*{8}{c}}
\toprule
 & \multicolumn{4}{c}{Random split} & \multicolumn{4}{c}{Temporal split} \\
\cmidrule(lr){2-5}\cmidrule(lr){6-9}
 & \multicolumn{2}{c}{F1} & \multicolumn{2}{c}{AUROC} & \multicolumn{2}{c}{F1} & \multicolumn{2}{c}{AUROC} \\
\cmidrule(lr){2-3}\cmidrule(lr){4-5}\cmidrule(lr){6-7}\cmidrule(lr){8-9}
Context length & Full & Head & Full & Head & Full & Head & Full & Head \\
\midrule
\multicolumn{9}{l}{\textit{BERT}} \\
32 (Title) & 59.98 & 42.27 & 64.94 & 51.36 & 58.46 & 32.43 & 73.79 & 49.81 \\
128 & 66.67 & 53.93 & 75.29 & 57.02 & 69.43 & 58.32 & 80.69 & 62.37 \\
256 & 73.75 & 57.96 & 78.86 & 59.45 & 70.43 & 58.97 & 83.92 & 66.93 \\
512 & 73.67 & 60.65 & 81.07 & 64.86 & 71.16 & 57.89 & 84.37 & 68.50 \\
\addlinespace
\multicolumn{9}{l}{\textit{Longformer}} \\
32 (Title) & 57.14 & 55.18 & 64.62 & 60.75 & 62.14 & 56.94 & 76.34 & 68.47 \\
128 & 68.18 & 67.22 & 76.66 & 71.80 & 67.94 & 66.09 & 80.98 & 77.17 \\
256 & 70.10 & 65.99 & 78.06 & 73.10 & 70.82 & 69.12 & 83.99 & 79.96 \\
512 & 73.03 & 64.93 & 81.61 & 66.76 & 73.00 & 63.57 & 86.38 & 74.07 \\
1024 & 75.00 & 62.79 & 83.70 & 68.30 & 74.56 & 62.55 & 88.81 & 72.15 \\
2048 & 76.65 & 62.50 & 84.96 & 67.79 & 77.07 & 61.91 & 89.80 & 70.78 \\
\bottomrule
\end{tabular}

    \begin{tablenotes}[flushleft]
        \footnotesize \item Notes: Test-set F1 and AUROC, in percent, at Youden-optimized thresholds. ``Full'' fine-tunes the backbone and classification head; ``Head'' freezes the backbone and trains only the classification head.
    \end{tablenotes}
    \end{threeparttable}
    }%
    \label{tab:head_vs_full}
\end{table}

\begin{figure}[!htbp]
    \centering
    \includegraphics[width=0.85\textwidth]{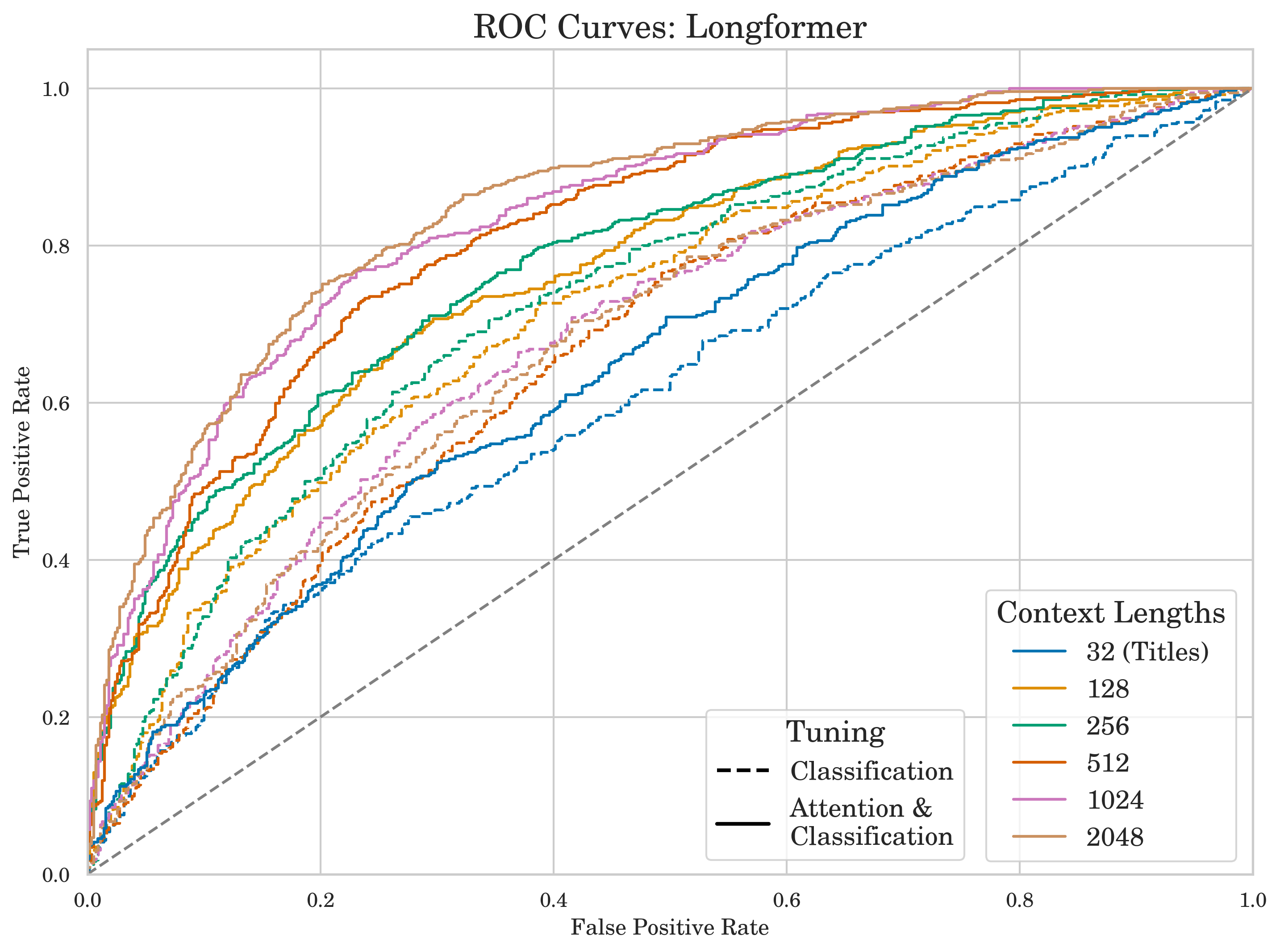}
    \caption{Longformer ROC Curves: Full Fine-Tuning versus Classification Head Only}
    \label{fig:roc_head}
    \fignote{Random split, held-out test set. Solid lines fine-tune the backbone and classification head (``Attention \& Classification''); dashed lines train only the classification head (``Classification''). Colors denote context length.}
\end{figure}

\clearpage

\subsection{Auditor Certainty and Article Length}

Figure~\ref{fig:misclass} complements Figure 1B of the main text by grouping held-out test articles according to auditor-reported certainty. The BOW misclassification rate is 42.39 percent among articles auditors were unsure about and 38.10 percent among those they marked certain. The corresponding rates for Longformer (2048) are 36.96 and 18.13 percent, so the reduction in error is concentrated among articles that auditors judged unambiguous.

Figure~\ref{fig:length} splits the random-split test set by whether an article fits within the model's context window. At 256 tokens, the 986 articles that exceed the window have an F1 of 0.68, against 0.81 for the 54 that fit. The gap shrinks to between 2 and 5 points at 1024 and 2048 tokens, where most articles fit entirely. Because article length is correlated with content, these comparisons are descriptive; they indicate that longer windows recover information that truncation discards.

\begin{figure}[!htbp]
    \centering
    \includegraphics[width=0.7\textwidth]{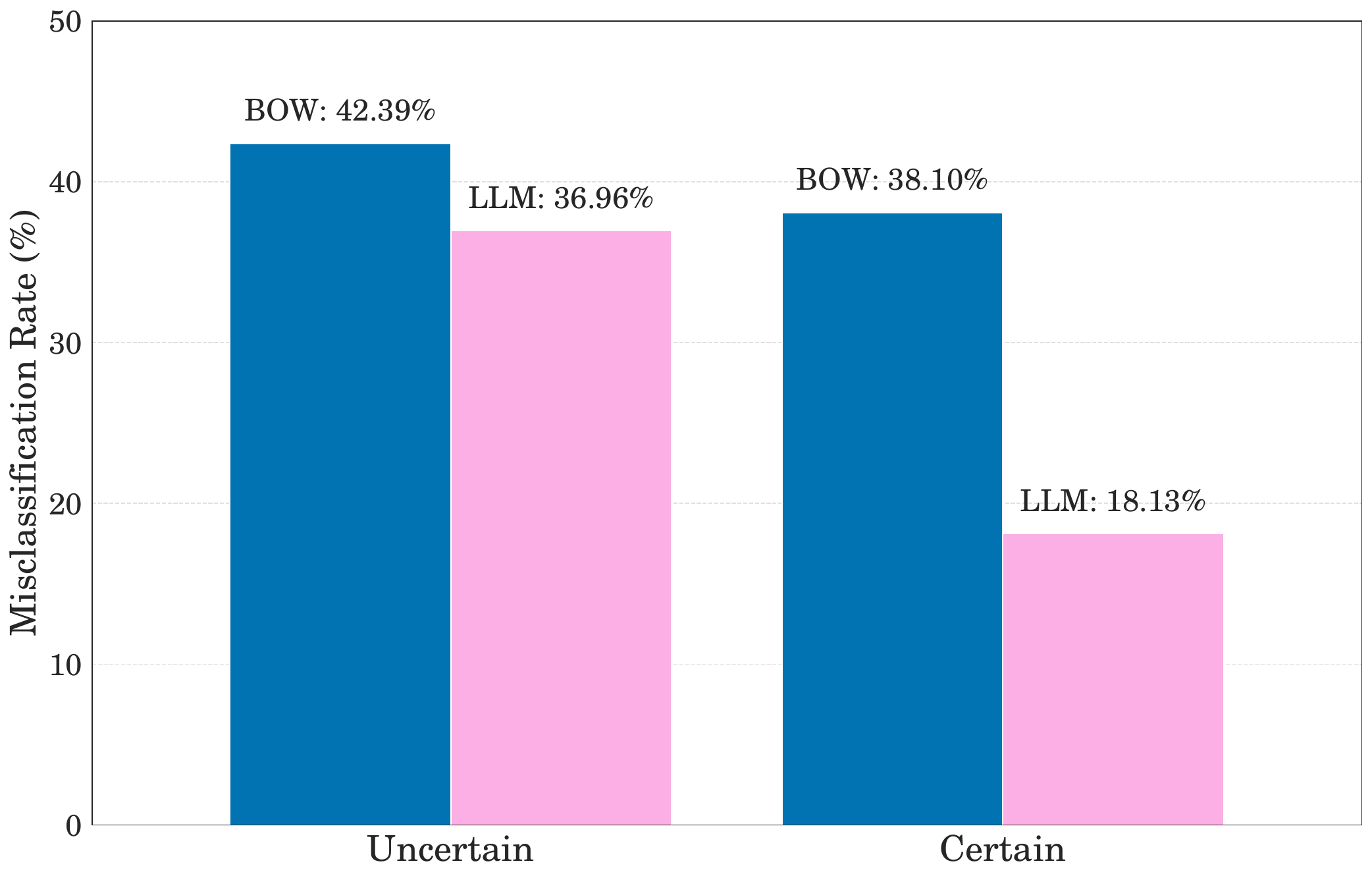}
    \caption{Misclassification Rate by Auditor Certainty}
    \label{fig:misclass}
    \fignote{Share of held-out test articles misclassified by BOW and Longformer (2048) at the Youden threshold. ``Uncertain'' pools auditor certainty levels 1 and 2 (184 articles); ``Certain'' is level 3 (546 articles). Articles coded with certainty level 4 are excluded, as in Figure 1B of the main text.}
\end{figure}

\begin{figure}[!htbp]
    \centering
    \includegraphics[width=0.7\textwidth]{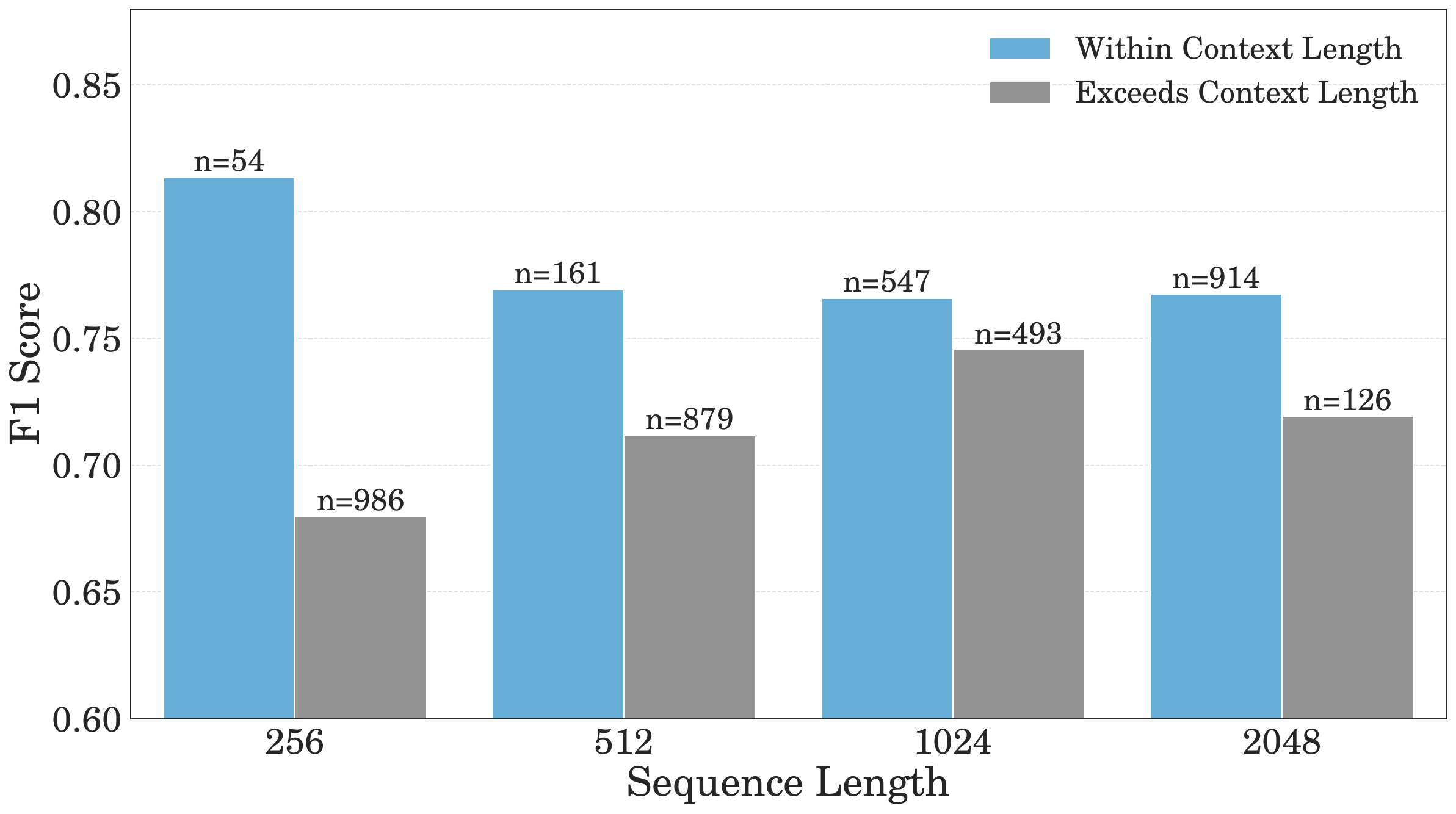}
    \caption{F1 by Context Length for Articles Within and Beyond the Context Window}
    \label{fig:length}
    \fignote{Random-split Longformer models evaluated on the 1{,}040 held-out test articles. Within each context length, articles are grouped by whether their tokenized length fits within the window; labels report the number of articles in each group.}
\end{figure}

\clearpage

\section{Multi-label Policy Classification}\label{sec:topics}
The BBD audit also records which policy categories each article concerns. Auditors could assign any combination of 16 categories, but category labels can be reliably linked for only a subset of articles and are heavily imbalanced. We focus on the seven categories with established BOW dictionaries: monetary policy; fiscal policy and government spending; tax policy; entitlement programs; healthcare policy; financial regulation; and trade policy. We select all 2{,}973 articles with at least one positive category label and add a random sample of 1{,}000 articles with none, for 3{,}973 articles in total. We split these 70/20/10 into training, validation, and test sets using the iterative stratification procedure of \citet{strat_sample}, which preserves each category's share across partitions. Table~\ref{tab:topic_split} reports the resulting counts for all ten categories coded in this sample.

Assigning articles to policy categories is a multi-label problem in which each article can belong to up to $C$ categories at once. To account for class imbalance, we weight equation~(\ref{eq:bce}) by the inverse proportion of positive labels in each category. Let $w_c$ denote the inverse positive rate for category $c \in \{1,\dots,C\}$, $y_{i,c}$ the auditor label of article $i$ for category $c$, and $p_{i,c}$ the corresponding predicted probability. The weighted loss is
\begin{align}
    \mathrm{BCE}_{w}
    &= - \frac{1}{N} \sum_{c=1}^C w_c \sum_{i=1}^N
    \Big( y_{i,c} \log(p_{i,c}) + \big(1 - y_{i,c}\big) \log\big(1-p_{i,c}\big) \Big), \label{eq:bce_w}
\end{align}
which equalizes each category's contribution to the loss regardless of its prevalence. The BOW benchmarks use the category term sets established by BBD, supplemented in some categories by updates hosted on policyuncertainty.com (Table~\ref{tab:keywords}).

Tables~\ref{tab:category_perf_youden} and \ref{tab:category_perf_f1} report accuracy and F1 by category at Youden and F1-optimized thresholds. BOW and SVM produce binary labels, so their entries do not depend on the threshold rule. Because positive labels are rare, accuracy rewards classifiers that seldom predict a category, and F1 is the more informative comparison. Figure~\ref{fig:category_f1} plots F1 for the four categories with more than 100 positive labels. At F1-optimized thresholds, a single multi-label Longformer (2048) exceeds BOW F1 by 40 to 132 percent in these categories and exceeds the SVM in all four. Estimates for the three smaller categories rest on five to seven positive test articles each and are correspondingly imprecise.

\begin{table}[!htbp]
    \caption{Policy Category Counts by Data Split}
    \centering
    \resizebox{0.7\textwidth}{!}{%
    \begin{threeparttable}
    \begin{tabular}{lccc}
\toprule
Topic & Train & Validation & Test \\
\midrule
Total articles & 2,781 & 835 & 357 \\
No topic label & 696 & 211 & 93 \\
\midrule
Tax Policy & 358 & 107 & 46 \\
Trade Policy & 322 & 97 & 41 \\
Monetary Policy & 245 & 74 & 31 \\
Healthcare Policy & 42 & 13 & 5 \\
Competition Policy & 73 & 22 & 10 \\
Financial Regulation & 52 & 15 & 7 \\
Entitlement Programs & 44 & 13 & 6 \\
Energy and Environment & 157 & 47 & 20 \\
Fiscal Policy \& Govt Spending & 376 & 113 & 48 \\
Political Conflict \& Leadership Change & 524 & 158 & 67 \\
\bottomrule
\end{tabular}

    \begin{tablenotes}[flushleft]
        \footnotesize \item Notes: ``No topic label'' indicates articles without any category assignment. Category counts exceed the number of articles because articles can be assigned to multiple categories. The classifiers predict the seven categories listed in Table~\ref{tab:keywords}.
    \end{tablenotes}
    \end{threeparttable}
    }%
    \label{tab:topic_split}
\end{table}

\begin{table}[!htbp]
    \centering
    \begin{threeparttable}
        \caption{Policy Category Keyword Dictionaries}
        \label{tab:keywords}
        \small
        \begin{tabularx}{\textwidth}{lX}
            \toprule
            Category & Keywords \\
            \midrule
            Monetary Policy & federal reserve, the fed, money supply, open market operations, quantitative easing, monetary policy, fed funds rate, overnight lending rate, bernanke, volcker, greenspan, central bank, interest rates, fed chairman, fed chair, lender of last resort, discount window, european central bank, ecb, bank of england, bank of japan, boj, bank of china, bundesbank, bank of france, bank of italy \\
            \addlinespace
            Fiscal Policy \& Govt Spending & taxes, tax, taxation, taxed, government spending, federal budget, budget battle, balanced budget, defense spending, military spending, entitlement spending, fiscal stimulus, budget deficit, federal debt, national debt, Gramm-Rudman, debt ceiling, fiscal footing, government deficits, balance the budget \\
            \addlinespace
            Tax Policy & taxes, tax, taxation, taxed \\
            \addlinespace
            Entitlement Programs & entitlement program, entitlement spending, government entitlements, social security, medicaid, medicare, government welfare, welfare reform, unemployment insurance, unemployment benefits, food stamps, afdc, tanf, wic program, disability insurance, part d, oasdi, supplemental nutrition assistance program, earned income tax credit, eitc, head start program, public assistance, government subsidized housing \\
            \addlinespace
            Healthcare Policy & health care, medicaid, medicare, health insurance, malpractice tort reform, malpractice reform, prescription drugs, drug policy, food and drug administration, fda, medical malpractice, prescription drug act, medical insurance reform, medical liability, part d, affordable care act, obamacare \\
            \addlinespace
            Financial Regulation & banking supervision, glass-steagall, tarp, bank supervision, thrift supervision, dodd-frank, financial reform, commodity futures trading commission, cftc, house financial services committee, basel, capital requirement, volcker rule, bank stress test, securities and exchange commission, sec, deposit insurance, fdic, fslic, ots, occ, firrea, truth in lending \\
            \addlinespace
            Trade Policy & import tariffs, import duty, import barrier, government subsidies, government subsidy, wto, world trade organization, trade treaty, trade agreement, trade policy, trade act, doha round, uruguay round, gatt, dumping \\
            \bottomrule
        \end{tabularx}
        \begin{tablenotes}[flushleft]
            \footnotesize
            \item Notes: Term sets used to construct the BOW category classifiers for the multi-label exercises.
        \end{tablenotes}
    \end{threeparttable}
\end{table}

\begin{table}[!htbp]
    \caption{Multi-label Policy Classification Performance (Youden Thresholds)}
    \centering
    \resizebox{\textwidth}{!}{%
    \begin{threeparttable}
    \begin{tabular}{lc*{6}{c}}
\toprule
 & & \multicolumn{3}{c}{Accuracy} & \multicolumn{3}{c}{F1 score} \\
\cmidrule(lr){3-5}\cmidrule(lr){6-8}
Policy category & Positive articles & BOW & SVM & LLM & BOW & SVM & LLM \\
\midrule
Tax Policy & 511 & 75.38 & 89.64 & 87.39 & 49.95 & 50.67 & 60.87 \\
Trade Policy & 460 & 87.36 & 91.32 & 87.68 & 29.10 & 55.07 & 59.26 \\
Monetary Policy & 350 & 71.76 & 92.44 & 86.83 & 33.21 & 47.06 & 51.55 \\
Healthcare Policy & 60 & 96.30 & 99.44 & 98.32 & 40.49 & 75.00 & 62.50 \\
Financial Regulation & 74 & 31.97 & 98.32 & 89.64 & 4.11 & 25.00 & 27.45 \\
Entitlement Programs & 63 & 93.51 & 98.60 & 93.00 & 21.34 & 28.57 & 32.43 \\
Fiscal Policy \& Govt Spending & 537 & 63.13 & 87.11 & 78.99 & 30.93 & 45.24 & 52.23 \\
\bottomrule
\end{tabular}

    \begin{tablenotes}[flushleft]
        \footnotesize \item Notes: Accuracy and F1 score, in percent, by policy category. LLM thresholds are selected separately for each category by Youden's index on the validation set. Positive articles are counts across the full labeled sample.
    \end{tablenotes}
    \end{threeparttable}
    }%
    \label{tab:category_perf_youden}
\end{table}

\begin{table}[!htbp]
    \caption{Multi-label Policy Classification Performance (F1-Optimized Thresholds)}
    \centering
    \resizebox{\textwidth}{!}{%
    \begin{threeparttable}
    \begin{tabular}{lc*{6}{c}}
\toprule
 & & \multicolumn{3}{c}{Accuracy} & \multicolumn{3}{c}{F1 score} \\
\cmidrule(lr){3-5}\cmidrule(lr){6-8}
Policy category & Positive articles & BOW & SVM & LLM & BOW & SVM & LLM \\
\midrule
Tax Policy & 511 & 75.38 & 89.64 & 92.72 & 49.95 & 50.67 & 69.77 \\
Trade Policy & 460 & 87.36 & 91.32 & 93.84 & 29.10 & 55.07 & 67.65 \\
Monetary Policy & 350 & 71.76 & 92.44 & 94.12 & 33.21 & 47.06 & 66.67 \\
Healthcare Policy & 60 & 96.30 & 99.44 & 99.72 & 40.49 & 75.00 & 90.91 \\
Financial Regulation & 74 & 31.97 & 98.32 & 97.48 & 4.11 & 25.00 & 30.77 \\
Entitlement Programs & 63 & 93.51 & 98.60 & 98.88 & 21.34 & 28.57 & 66.67 \\
Fiscal Policy \& Govt Spending & 537 & 63.13 & 87.11 & 91.32 & 30.93 & 45.24 & 59.74 \\
\bottomrule
\end{tabular}

    \begin{tablenotes}[flushleft]
        \footnotesize \item Notes: Accuracy and F1 score, in percent, by policy category. LLM thresholds are selected separately for each category to maximize F1 on the validation set. BOW and SVM entries are identical to Table~\ref{tab:category_perf_youden}.
    \end{tablenotes}
    \end{threeparttable}
    }%
    \label{tab:category_perf_f1}
\end{table}

\begin{figure}[!htbp]
    \centering
    \includegraphics[width=\textwidth]{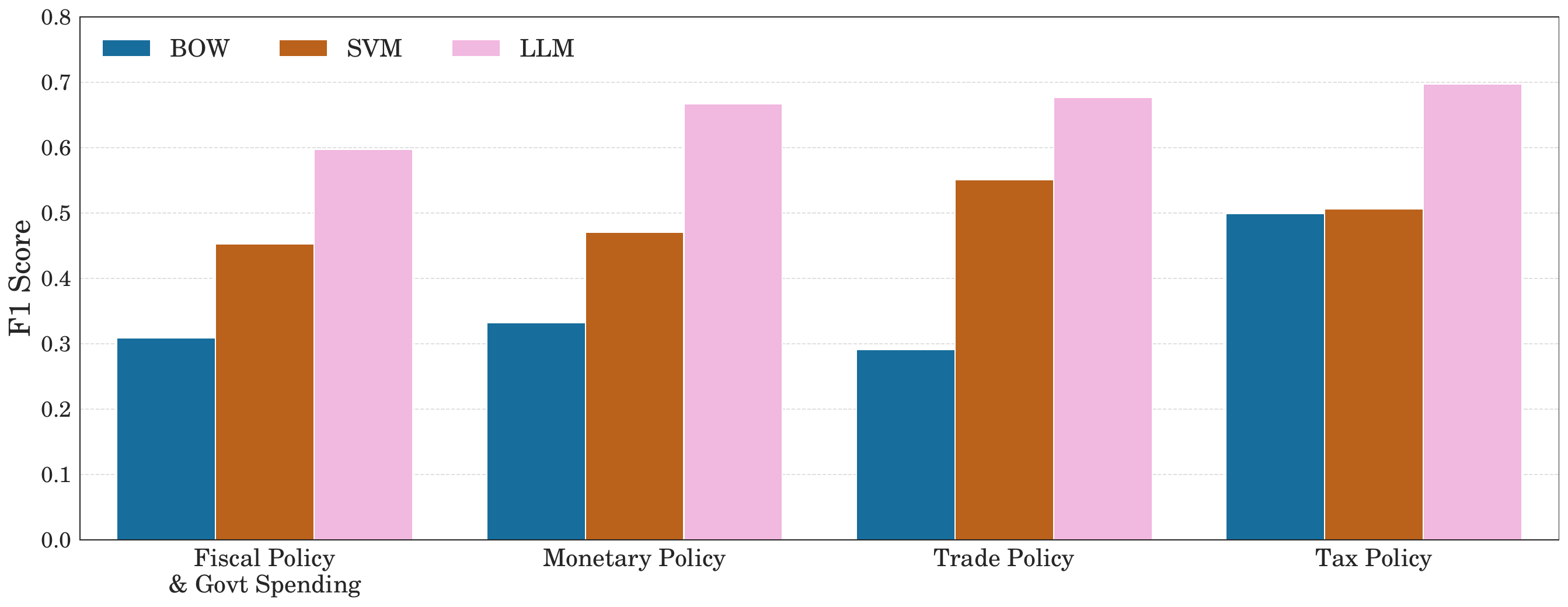}
    \caption{F1 by Policy Category}
    \label{fig:category_f1}
    \fignote{F1 scores for the four policy categories with more than 100 positive labels, using F1-optimized thresholds for the LLM (Table~\ref{tab:category_perf_f1}).}
\end{figure}

\clearpage

\section{Index Fidelity}\label{sec:fidelity}
Section III.A of the main text evaluates whether lower article-level error yields an index that tracks the human audit benchmark more closely. Tables~\ref{tab:audit_corr_models} and \ref{tab:audit_corr_thresholds} report the underlying correlations. Each index is built from the audit sample at quarterly frequency over 1985--2012 using equation (2) of the main text, with every audit article scored by the indicated classifier. The benchmark applies the same construction to the auditor labels $A_{it}$. Because the temporal split holds out 2006--2012, the post-2006 column is the clean out-of-sample comparison for temporal-split models.

Every LLM-based index tracks the audit series more closely than the BOW index in every period, and within each model and split the probabilistic index correlates at least as closely as the binary index. Under the temporal split, correlations decline modestly in the held-out period, from 0.77 to 0.70 for the binary Longformer index and from 0.84 to 0.78 for the probabilistic index. Table~\ref{tab:audit_corr_thresholds} holds Longformer (2048) fixed and varies the aggregation rule, as in Figure 2B of the main text. Over the full period and under both splits, probabilistic aggregation correlates most closely with the audit series and the precision-optimized threshold, which admits the fewest articles, correlates least.

\begin{table}[!htbp]
    \caption{Correlation with the Audit Series by Model and Sample Split}
    \centering
    \resizebox{0.7\textwidth}{!}{%
    \begin{threeparttable}
    \begin{tabular}{lccc}
\toprule
 & All years & Pre-2006 & Post-2006 \\
\midrule
BOW & 0.33 & 0.48 & 0.24 \\
Longformer Binary (Temporal) & 0.74 & 0.77 & 0.70 \\
Longformer Prob (Temporal) & 0.81 & 0.84 & 0.78 \\
Longformer Binary (Random) & 0.79 & 0.79 & 0.80 \\
Longformer Prob (Random) & 0.82 & 0.82 & 0.82 \\
Llama Binary (Temporal) & 0.75 & 0.76 & 0.71 \\
Llama Prob (Temporal) & 0.80 & 0.83 & 0.71 \\
Llama Binary (Random) & 0.71 & 0.69 & 0.72 \\
Llama Prob (Random) & 0.78 & 0.76 & 0.80 \\
\bottomrule
\end{tabular}

    \begin{tablenotes}[flushleft]
        \footnotesize \item Notes: Correlation between the index constructed from auditor labels and each model-based index, quarterly, 1985--2012. Indices follow equation (2) of the main text across the eight audit newspapers, with standardization and normalization over years before 2011. Binary indices use Youden-optimized thresholds; ``Prob'' indices use probabilistic aggregation. ``Temporal'' and ``Random'' denote the sample split used to train the classifier.
    \end{tablenotes}
    \end{threeparttable}
    }%
    \label{tab:audit_corr_models}
\end{table}

\begin{table}[!htbp]
    \caption{Correlation with the Audit Series by Aggregation Rule}
    \centering
    \resizebox{0.6\textwidth}{!}{%
    \begin{threeparttable}
    \begin{tabular}{lccc}
\toprule
 & All years & Pre-2006 & Post-2006 \\
\midrule
\multicolumn{4}{l}{\textit{Panel A. Random split}} \\
Probabilistic & 0.82 & 0.82 & 0.82 \\
Youden & 0.79 & 0.79 & 0.80 \\
Recall & 0.77 & 0.79 & 0.75 \\
Precision & 0.70 & 0.71 & 0.75 \\
\addlinespace
\multicolumn{4}{l}{\textit{Panel B. Temporal split}} \\
Probabilistic & 0.81 & 0.84 & 0.78 \\
Youden & 0.74 & 0.77 & 0.70 \\
Recall & 0.76 & 0.80 & 0.72 \\
Precision & 0.70 & 0.70 & 0.71 \\
\bottomrule
\end{tabular}

    \begin{tablenotes}[flushleft]
        \footnotesize \item Notes: Correlation between the index constructed from auditor labels and Longformer (2048) indices built with probabilistic aggregation and with the Youden-, recall-, and precision-optimized thresholds, quarterly, 1985--2012. Thresholds are selected on each split's validation set. Construction otherwise follows Table~\ref{tab:audit_corr_models}.
    \end{tablenotes}
    \end{threeparttable}
    }%
    \label{tab:audit_corr_thresholds}
\end{table}

\clearpage

\section{Historical Index}\label{sec:historical}
We construct the historical index by scoring each American Stories article \citep{dell_newspapers} with Longformer (2048) and aggregating by newspaper and month following equations (1) and (2) of the main text. Figure~\ref{fig:as_counts} reports monthly article counts. Coverage in American Stories varies widely across newspapers and over time, and many newspapers appear in only a handful of months. We therefore exclude newspapers that contribute more than one article in fewer than two months; in the 1900--1950 sample used in Section IV of the main text, this rule retains 1{,}598 of 1{,}655 newspapers. Because the archive ends in 1964, the normalization window $T_0$ spans the full sample: each newspaper's share is scaled by its standard deviation over the months in which it appears, and the cross-newspaper average is rescaled to a mean of 100.

Figure~\ref{fig:as_vs_bbd} plots the precision-threshold index against the published BBD historical index over their 1900--1964 overlap, where the two series correlate at 0.33. Figure~\ref{fig:as_thresholds} plots the full 1800--1964 series under all four constructions, and Table~\ref{tab:as_corr} reports their pairwise correlations. The three binary indices move closely together, with correlations between 0.84 and 0.98, while the probabilistic index correlates less closely with each, consistent with the dilution under probabilistic aggregation discussed in Section III.A of the main text.

Coverage thins at both ends of the sample. In the earliest months and in late 1964, few newspapers contribute in a given month, and binary indices at strict thresholds can move between zero and several hundred from one month to the next. We retain these months in all reported statistics, including the correlation with the BBD index, and individual spikes at the tails of the series should be interpreted with caution.

\begin{figure}[!htbp]
    \centering
    \includegraphics[width=\textwidth]{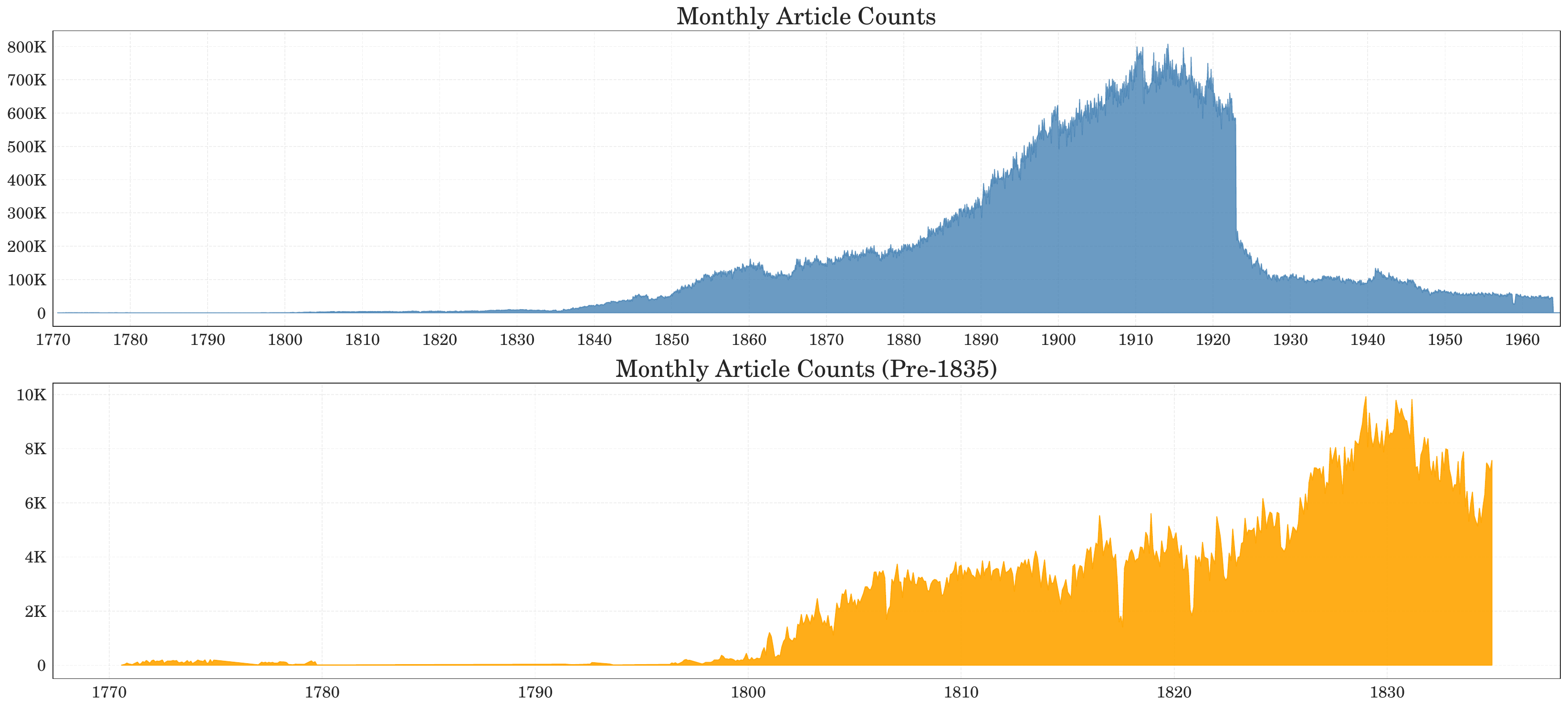}
    \caption{American Stories Monthly Article Counts}
    \label{fig:as_counts}
    \fignote{Number of digitized newspaper articles per month in American Stories. The bottom panel shows the period before 1835. The index uses articles from 1800--1964.}
\end{figure}

\begin{figure}[!htbp]
    \centering
    \includegraphics[width=\textwidth]{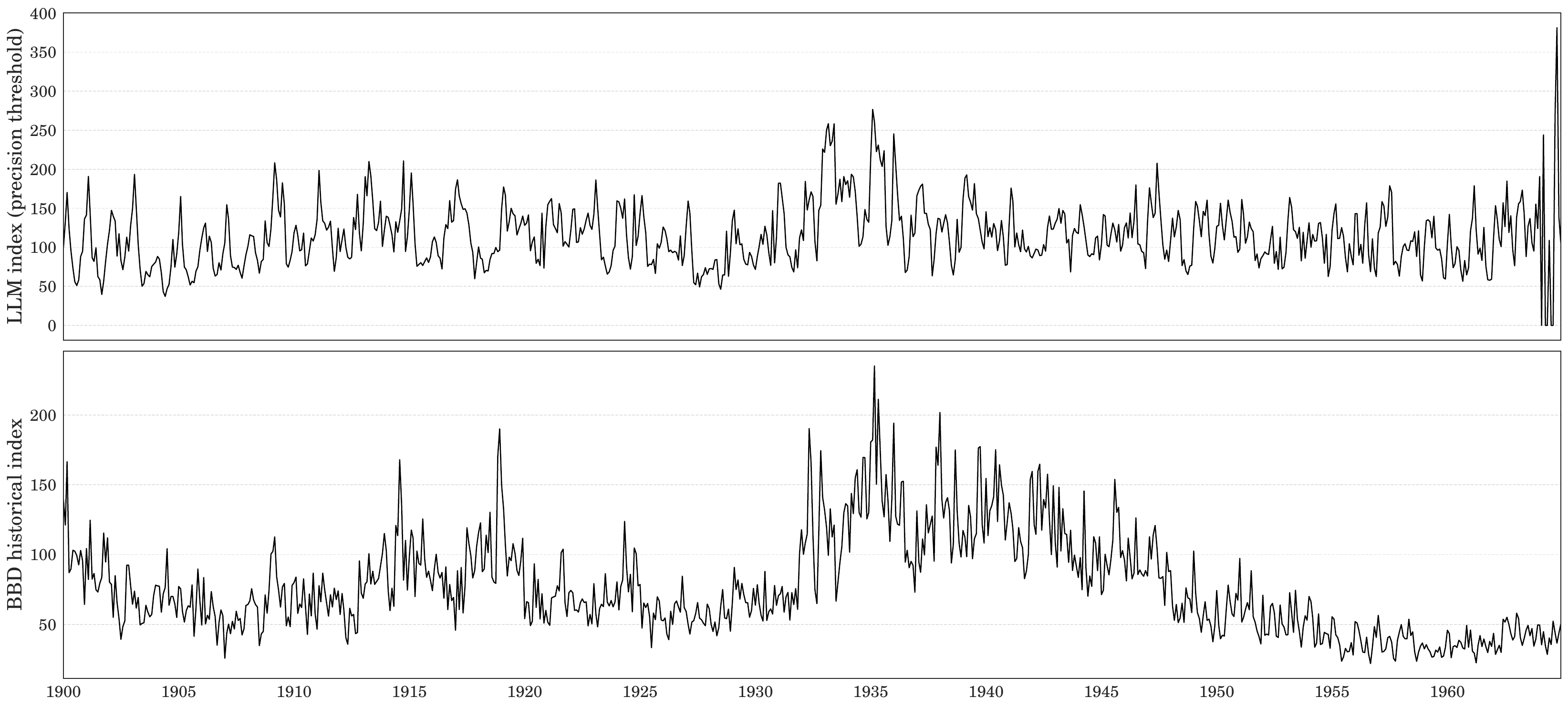}
    \caption{LLM-Based American Stories Index and the BBD Historical Index, 1900--1964}
    \label{fig:as_vs_bbd}
    \fignote{The top panel shows the monthly American Stories index built with Longformer (2048) at the precision-optimized threshold ($\tau = 0.74$). The bottom panel shows the published BBD historical index. The correlation between the two series over the 780 overlapping months is 0.33.}
\end{figure}

\begin{figure}[!htbp]
    \centering
    \includegraphics[width=\textwidth]{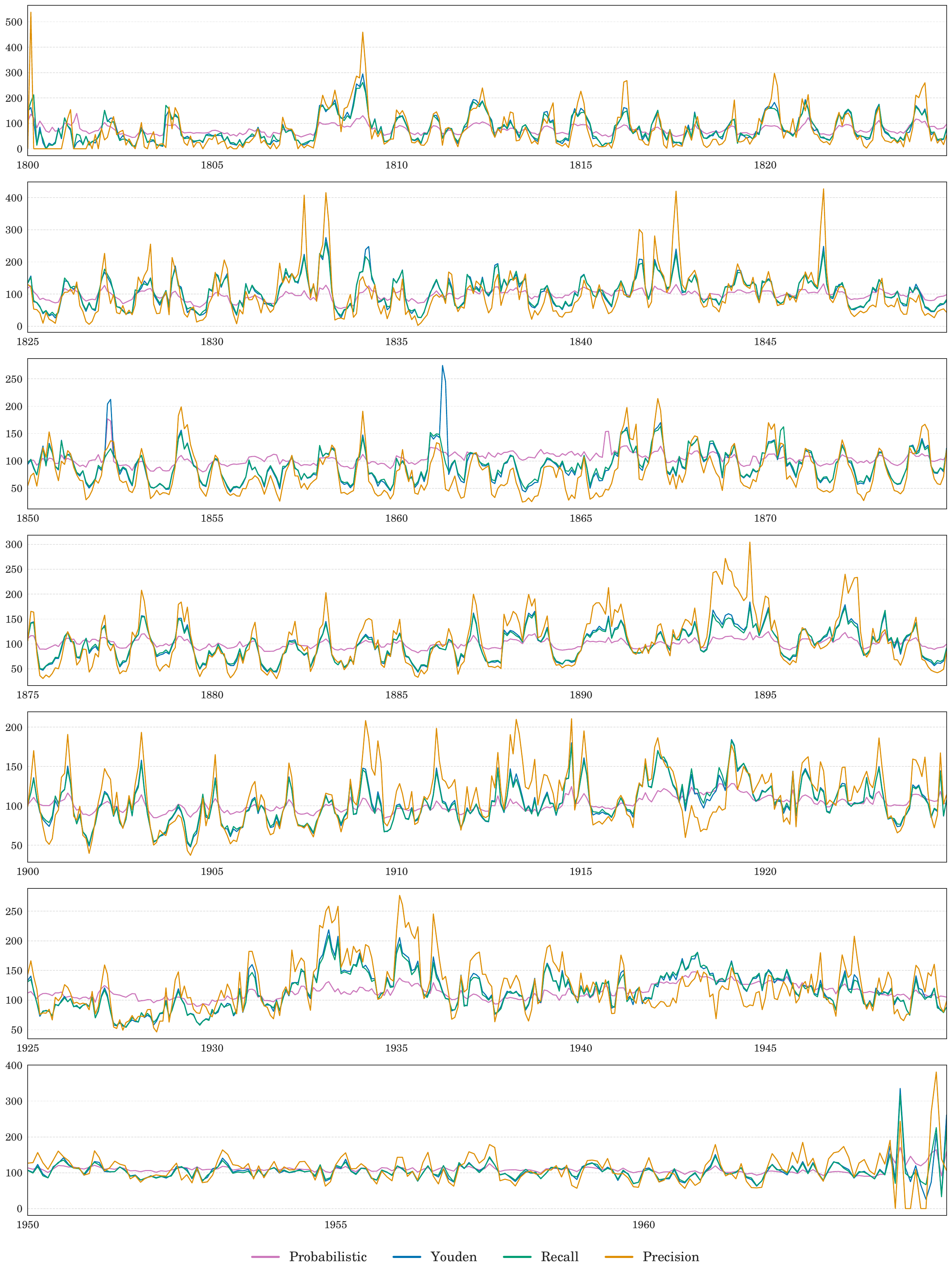}
    \caption{American Stories Index under Alternative Constructions, 1800--1964}
    \label{fig:as_thresholds}
    \fignote{Monthly American Stories indices built with Longformer (2048) using probabilistic aggregation and the Youden- ($\tau = 0.45$), recall- ($\tau = 0.40$), and precision-optimized ($\tau = 0.74$) thresholds. Each panel covers 25 years; the final panel covers 1950--1964.}
\end{figure}

\begin{table}[!htbp]
    \caption{Correlation Matrix of American Stories Indices}
    \centering
    \resizebox{0.7\textwidth}{!}{%
    \begin{threeparttable}
    \begin{tabular}{lcccc}
\toprule
 & Probabilistic & Youden & Recall & Precision \\
\midrule
Probabilistic & 1.00 & 0.73 & 0.74 & 0.58 \\
Youden & 0.73 & 1.00 & 0.98 & 0.85 \\
Recall & 0.74 & 0.98 & 1.00 & 0.84 \\
Precision & 0.58 & 0.85 & 0.84 & 1.00 \\
\bottomrule
\end{tabular}

    \begin{tablenotes}[flushleft]
        \footnotesize \item Notes: Pairwise correlations between the four monthly American Stories indices in Figure~\ref{fig:as_thresholds}, computed over the full sample period, 1800--1964.
    \end{tablenotes}
    \end{threeparttable}
    }%
    \label{tab:as_corr}
\end{table}

\clearpage

\section{Multilingual Extension}\label{sec:multilingual}
Llama 3.1 (8B) is pre-trained on text in many languages, so the classifier we fine-tune on English audit articles can be applied directly to non-English text. This section reports an exploratory application. The policy uncertainty literature includes many national and cross-country indices, typically built from keyword rules adapted or translated to each setting \citep{hites, hong, diakonova}. Aggregating indices built from different dictionaries, translations, and outlets combines differences in policy uncertainty with differences in measurement, and keyword methods are least developed for smaller languages. A single multilingual classifier offers a way to apply one definition of EPU across languages, building on recent progress in representation learning for low-resource languages \citep{costajussa}.

Two questions arise. The first is mechanical: holding content fixed, can the model process otherwise identical articles written in different languages? The second is substantive: does a classifier trained on English articles from U.S. newspapers capture how policy uncertainty is reported in other countries? Answering the second would require audit labels from native speakers across many languages, which we leave to future work. We address the first through a controlled translation exercise and then construct two illustrative indices.

We translate the validation and held-out test sets of the audit corpus into 29 languages\footnote{Translation uses the \texttt{deep-translator} Python package with the Google Translate backend.} and score every translated article with the fine-tuned Llama 3.1 (8B) classifier. Score distributions shift across languages (Figure~\ref{fig:language_dists}), so we re-optimize the decision threshold on each language's translated validation set. Figure~\ref{fig:multilingual}A reports bootstrapped test-set F1 by language at F1-optimized thresholds. Performance is highest for high-resource languages such as English and Spanish and lowest for languages less represented in pre-training, such as Tigrinya and Khmer; however, every language exceeds the F1 of the English BOW classifier applied to the original articles. Because the translated articles are U.S. newspaper coverage rendered in other languages rather than native journalism, this exercise tests the model's ability to process non-English input, not whether it captures how policy uncertainty is reported in each country.

To move from translations to native-language text, we construct a provisional Africa-wide index from Voice of America (VOA), a U.S.-funded international broadcaster that publishes news in many languages and maintains a public archive \citep{palenmichel}. We score articles in five languages, build country-level monthly series, and aggregate them into a GDP-weighted index.\footnote{The five languages are tied to countries with sufficient GDP data: Nigeria, Rwanda, Zimbabwe, Somalia, and Ethiopia.} The corpus contains 132{,}869 articles (Figure~\ref{fig:voa_counts}), but the precision-optimized threshold from the audit setting flags only 21, far too few for a monthly index. We therefore use the Youden-optimized threshold, trading some precision for a larger pool of positive articles. Figure~\ref{fig:multilingual}B plots the resulting series. Event annotations are assigned ex post, the index may weight countries that receive more VOA coverage more heavily, and VOA reflects the perspective of an external outlet even when it reports in local languages. The series is therefore best read as a measure of how African policy developments are framed in that outlet.

For a higher-volume example based on domestic media, we also construct an index from approximately two million Bangla-language articles published by nine local outlets in Bangladesh \citep{saad} (Figure~\ref{fig:bangladesh_counts}). With abundant data, we use the precision-optimized threshold. The resulting series (Figure~\ref{fig:multilingual}C) rises around salient domestic policy and macroeconomic events, but we treat it as an illustration rather than a validated national index. Together, the two indices show how threshold choice adapts to data availability: a balanced threshold extracts a usable signal from thin multilingual coverage, while abundant local news supports a stricter threshold.

\begin{figure}[!htbp]
    \centering
    \includegraphics[width=\textwidth]{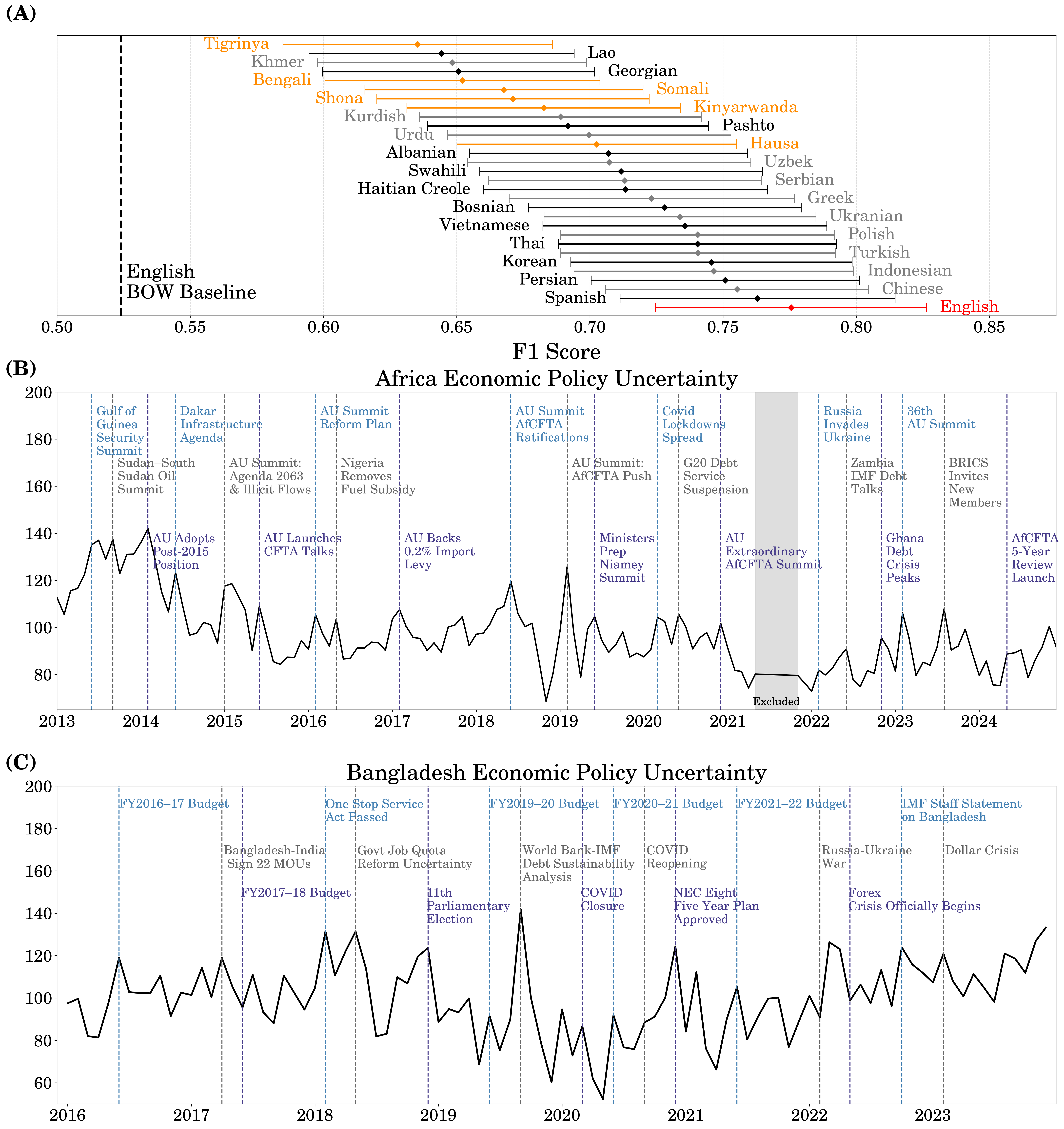}
    \caption{Multilingual Performance and Exploratory Cross-Country Indices}
    \label{fig:multilingual}
    \fignote{Panel A plots bootstrapped test-set F1 for the fine-tuned Llama 3.1 (8B) classifier on translations of the audit corpus into 29 languages, using language-specific F1-optimized thresholds. The dashed line marks the English BOW baseline; English is shown in red, and highlighted labels mark languages used in Panels B and C. Panel B shows a GDP-weighted Africa index built from VOA articles in five languages at the Youden-optimized threshold; the shaded interval is excluded. Panel C shows a Bangladesh index built from Bangla-language articles in nine local outlets at the precision-optimized threshold. Event annotations are assigned ex post.}
\end{figure}

\begin{figure}[!htbp]
    \centering
    \includegraphics[width=0.9\textwidth]{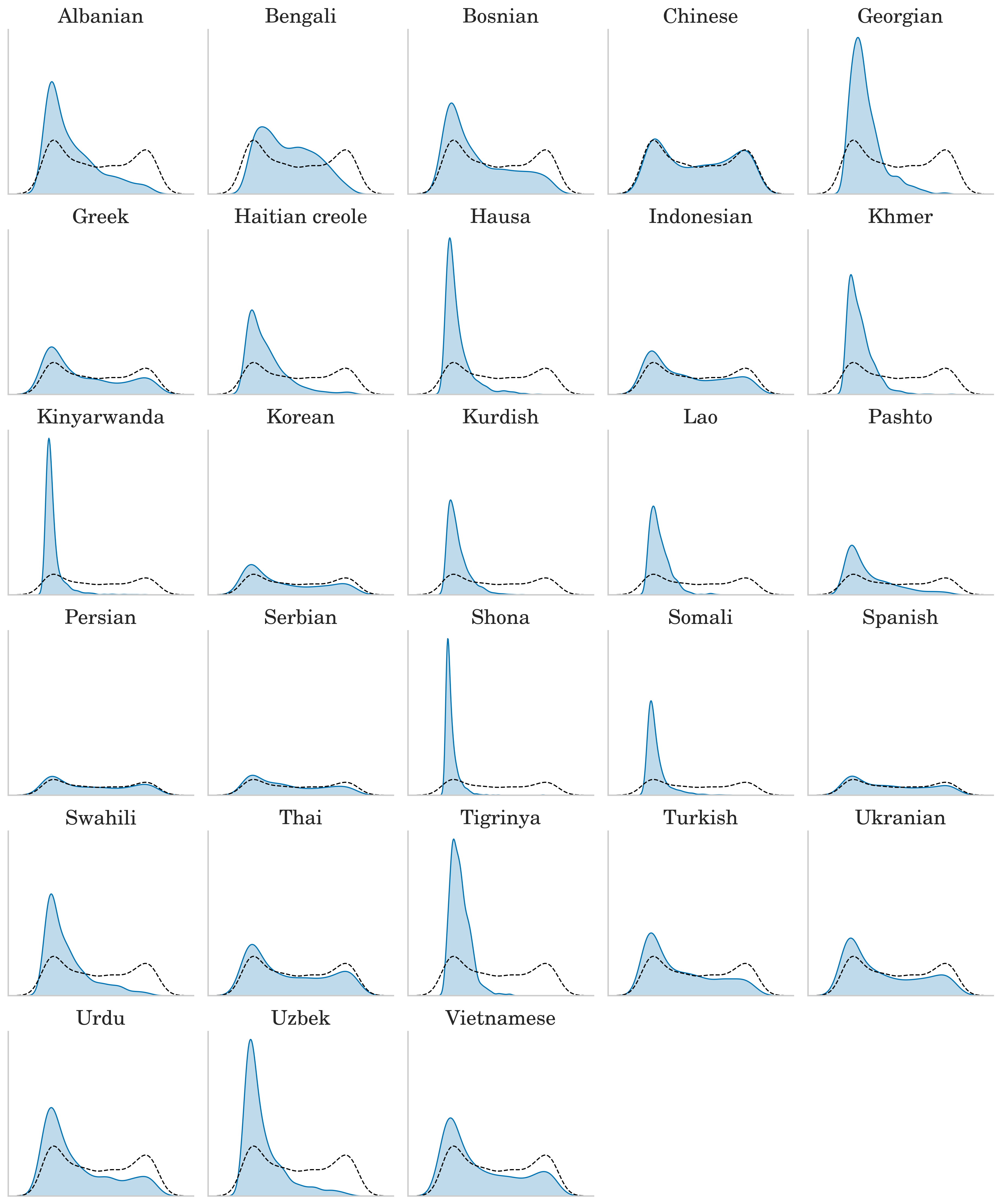}
    \caption{Distribution of Predicted Probabilities by Language}
    \label{fig:language_dists}
    \fignote{For each language, kernel density estimates compare the distribution of Llama 3.1 (8B) scores on English articles with the distribution on articles in the indicated language.}
\end{figure}

\begin{figure}[!htbp]
    \centering
    \includegraphics[width=\textwidth]{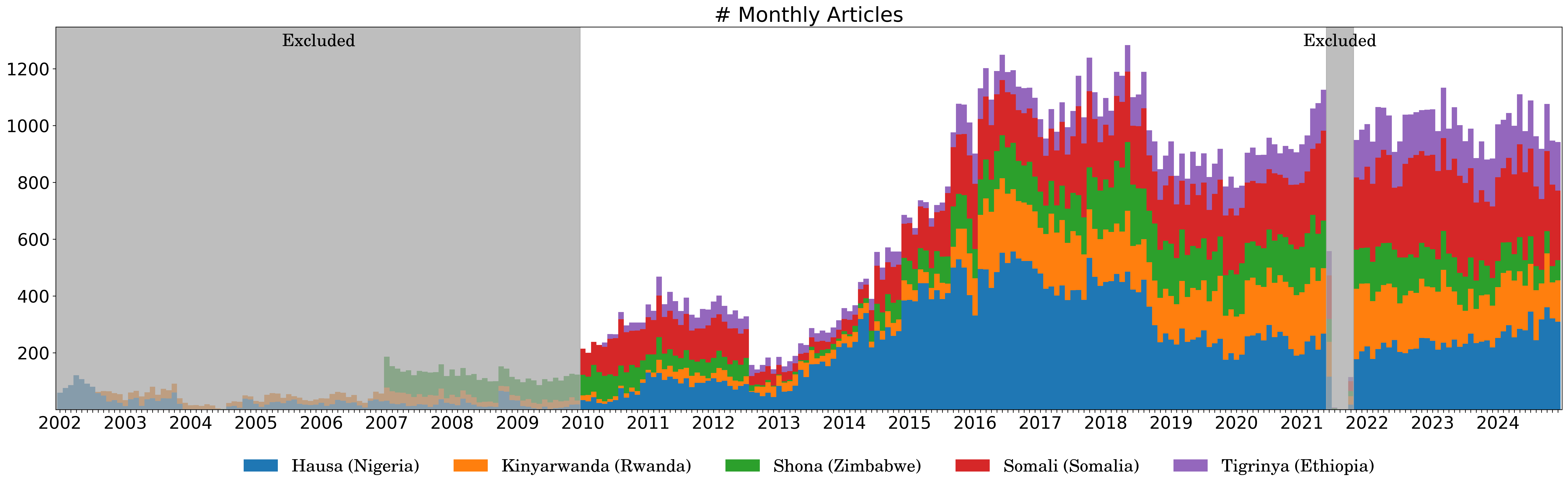}
    \caption{Voice of America Monthly Article Counts}
    \label{fig:voa_counts}
    \fignote{Number of VOA articles per month in the five languages used to construct the Africa index.}
\end{figure}

\begin{figure}[!htbp]
    \centering
    \includegraphics[width=\textwidth]{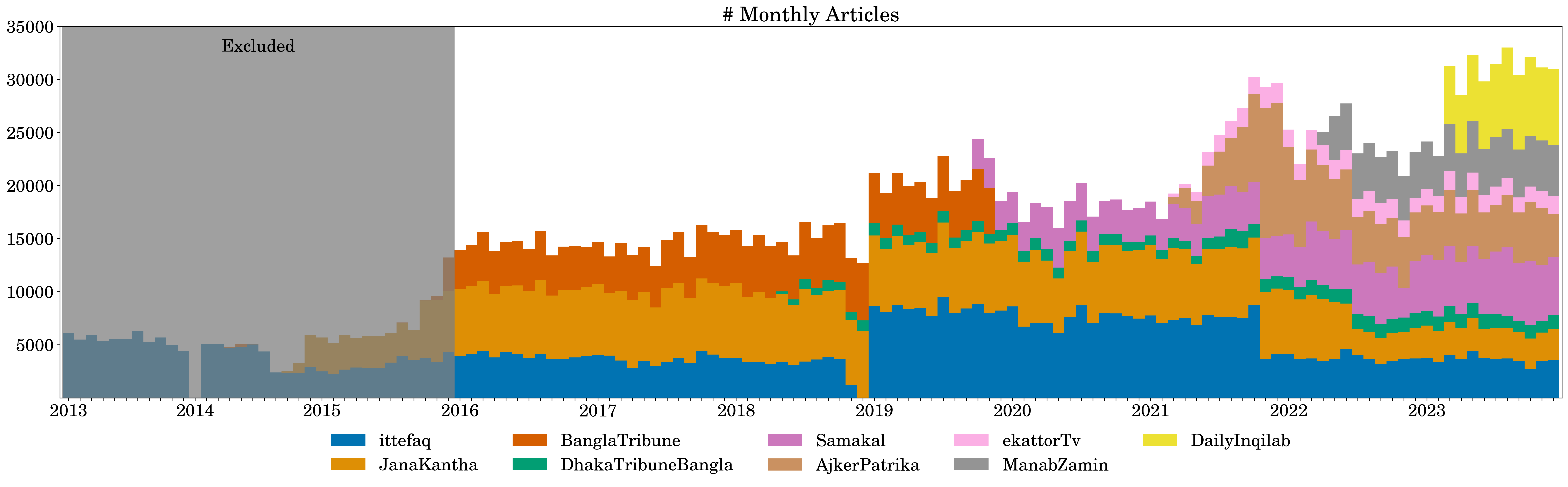}
    \caption{Bangladesh Monthly Article Counts}
    \label{fig:bangladesh_counts}
    \fignote{Number of articles per month across the nine Bangla-language outlets used to construct the Bangladesh index.}
\end{figure}

\clearpage

\section{Example Code}\label{sec:code}
Figure~\ref{fig:llm_code} illustrates the Hugging Face workflow used to fine-tune and score the classifiers. The example loads a pre-trained Longformer checkpoint and tokenizer, truncates each article to the chosen context length, pads batches dynamically, and passes the data to the \texttt{Trainer} API. Replacing the base checkpoint with one of our released models applies the fine-tuned classifier and returns article-level scores $\hat{p}_{ijt}$, which can then be thresholded and aggregated as in Definition 1 of the main text.

\begin{figure}[!htbp]
    \centering
    \includegraphics[width=0.85\textwidth]{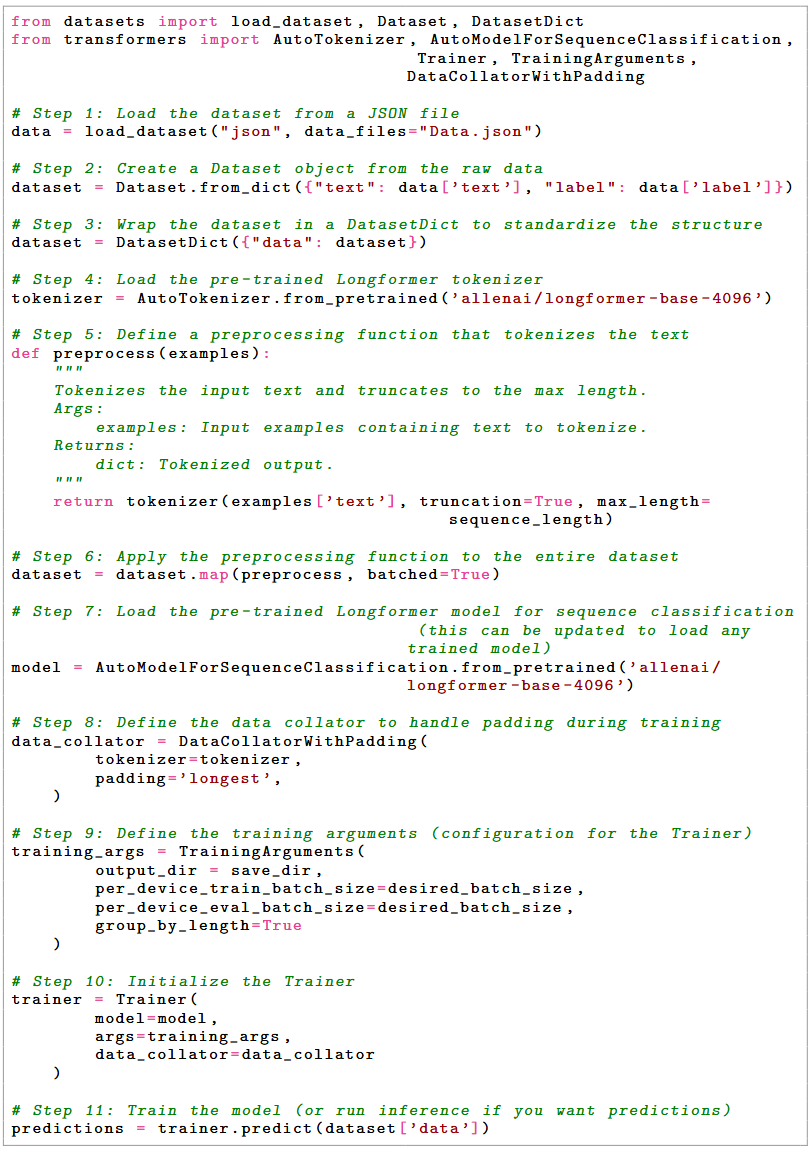}
    \caption{Hugging Face Implementation Example}
    \label{fig:llm_code}
    \fignote{Python example of the Hugging Face pipeline. The code loads the base \texttt{allenai/longformer-base-4096} checkpoint and tokenizer, tokenizes articles with truncation at the chosen sequence length, pads batches with a data collator, and runs the \texttt{Trainer} API. The final step calls \texttt{trainer.predict} to generate predictions; calling \texttt{trainer.train} instead fine-tunes the model.}
\end{figure}

\clearpage


\end{bibunit}

\end{document}